\documentclass{aa}

\usepackage{graphics}
\begin{document}

   \thesaurus{11     
              (11.17.3)}  
   \title{Quasar variability: correlations with amplitude}

   \author{M.R.S. Hawkins}

   \offprints{M.R.S. Hawkins}

   \institute{Institute for Astronomy (IfA), University of Edinburgh,
             Royal Observatory, Blackford Hill Edinburgh EH9 3HJ,
             Scotland\\
              email: mrsh@roe.ac.uk}

   \date{Received <date> / accepted <date>}

   \maketitle

   \begin{abstract}

The relation between quasar variability and parameters such as
luminosity and redshift has been a matter of hot debate over the last
few years with many papers on the subject.  Any correlations which can
be established will have a profound effect on models of quasar
structure and evolution.  The sample of quasars with redshifts in
ESO/SERC field 287 contains over 600 quasars in the range
$0 < z < 3.5$ and is now large enough to bin in luminosity and
redshift, and give definitive measures of the correlations.  We find
no significant correlation between amplitude and redshift, except
perhaps at very low redshift, but an inverse correlation between
amplitude and luminosity.  This is examined in the context of various
models for quasar variability.


   \end{abstract}

%

\section{Introduction}

One of the most important constraints on the structure of quasars is
variability.  Short term variations set limits on the size of the
emitting region, and differences in the nature of the variation in the
Xray, optical and radio domains give clues to the underlying structure.
Variations on longer timescales of a few years are not so easily
accounted for by the current black hole paradigm, and microlensing has
been put forward as an alternative explanation for the observations.
Although the most extensive monitoring of quasar variability has been
done in optical wavebands, there is still little consensus even over
the broad picture.  One of the first problems is to parametrise the
variation in such a way that it can be compared with models, or even
more general expectations.  The timescale of variation and the
amplitude are the two parameters which have been mostly studied,
although some authors have succeeded in confusing the two.  This
typically involves claiming that in a short run of data, objects
varying on a short timescale will achieve a larger amplitude than
those varying on a longer timescale, and so amplitude can be taken
as an (inverse) measure of timescale (Hook et al. \cite{hoo94}).\\

In this paper we shall confine our attention to the correlation of
amplitude with other parameters.  Amplitude is an easy parameter to
measure, and involves none of the difficulties inherent in estimating
timescale of variation.  There are modes of variability in which
some uncertainty will be introduced by the length of the run of
observations, but this is a problem which can arise in any time
series analysis. Most attention has so far been given to possible
correlations of amplitude with redshift or luminosity (eg Trevese
et al. \cite{tre89}, Giallongo et al. \cite{gia91},
Hook et al. \cite{hoo94}, Cristiani et al. \cite{cri96}), but even
in such straightforward situations there has been little agreement.\\

The reasons for the lack of consensus are not easy to understand,
but it is clear that any effect which is present is not large compared
with the cosmic scatter in the data, which in the case of amplitude
appears to be about one magnitude, much larger than any photometric
errors.  Perhaps the most likely cause of disagreement is in the
selection of samples for analysis.  In order to obtain a meaningful
measure of correlation it is essential that unbiassed samples are used,
and that they cover a large range of redshift and luminosity.  In this
paper particular attention is paid to the sample selection process, and
by extending the analysis to high redshifts the baseline for the
measurement of correlations is greatly extended over earlier samples.\\

\section{Quasar samples}

\subsection{Observational material}

The parent sample of quasars for the analysis of variability amplitude
comes from a large scale survey and monitoring programme being
undertaken in the ESO/SERC field 287, at 21h 28m, -45$^{\circ}$.
Quasars have been selected according to a number of criteria,
including colour, variability, radio emission and objective prism
spectra, or a combination of these techniques.  Redshifts for over 600
quasars have so far been obtained, and several complete samples
defined within specific limits of magnitude, redshift and position
on the sky (Hawkins \& V\'{e}ron \cite{haw95}).  A detailed description
of the survey is given by Hawkins \& V\'{e}ron (\cite{haw95}) and
Hawkins (\cite{haw96}).  Briefly, a large set of UK 1.2 Schmidt plates
spanning 20 years was scanned with the COSMOS and SuperCOSMOS measuring
machines to provide a catologue of some 200,000 objects in the central
19 square degrees of the plate.  These were calibrated with CCD frames
to provide light curves in $B_{J}$ and $R$, and colours in $UBVRI$.
Photometric errors have been discussed in detail in earlier papers
(Hawkins 1996 and references therein) and are $\sim 0.08$ magnitudes
for any individual machine measurement, and only weakly dependent on
magnitude.  There are approximately four plates in each year
which reduces the error to $\sim 0.04$ magnitudes per epoch.  This is
small compared with the amplitudes of interest in this paper, and
no attempt has been made to deconvolve it.

\subsection{Sample selection}

For the analysis in this paper three samples will be defined.
The first (UVX) is based only on position on the sky and ultraviolet
excess.  The area on the sky containing the sample is defined by a
number of AAT AUTOFIB fields and a 2dF field (Folkes et al. \cite{fol99}
and references therein), covering a total area of 7.0 square degrees.
Within this area all objects with $U-B < -0.2$ and $B_{J} < 21.0$ were
observed.  This was extended to $B_{J} < 21.5$ in the 2dF field.
The ultra-violet excess (UVX) cut although necessary to give a
relatively clean sample of quasar candidates, has the well known
limitation of only being effective for redshifts $z < 2.2$.  Beyond
this redshift quasars become red in $U-B$ as the Lyman forest enters
the $U$ band.  For samples at higher redshift, variability has
been found to provide a very useful criterion for quasar selection
(Hawkins \cite{haw96}).  The second sample for consideration herei
(VAR), was selected on this basis, with the requirement that the object
should lie in the same area of sky as for the first sample with a
magnitude limit $B_{J} < 21$, or anywhere in the measured area of the
plate with a magnitude limit $B_{J} < 19.5$, and should have an
amplitude $\delta m > 0.35$.  The defining epoch for the magnitudes
was the year 1977.  This sample was selected without any reference to
colour and so can be used to measure trends over a large range of
redshift ($z < 3.5$).  The third sample (AMP) was selected in a similar
manner, but over the whole measured area of the plate (19 square
degrees), and with an amplitude cut $\delta m > 1.1$.  There is some
evidence that these large amplitude objects form a distinct group,
which is discussed below.\\
 
The development of fibre fed spectrographs has meant that every
object included within a given set of search criteria can be
observed to give a high completeness level.  In the case of
fields observed with AUTOFIB, the existence of forbidden regions
in the 40 arcminute field, and the very variable throughput for
different fibres aligned at different positions on the spectrograph
slit meant that up to 20\% of spectra did not have sufficient signal
to provide an unambiguous redshift.  These objects where re-observed
later with the faint object spectrograph EFOSC on the ESO 3.6m
telescope at La Silla.  The 2dF observations where of very uniform
quality, and the redshift measures or other classification were almost
100\% successful from the fibre-feed spectra.\\

Details of all three samples, containing a total of 384 quasars,
are given in the Appendix.  $B$ magnitudes
are in the $B_{J}$ system defined by the IIIa-J emulsion and the GG395
filter, and refer to the year 1977.  The amplitude $\delta m$ is the
difference between maximum and minmum magnitude achieved over the 20
year run of data.  The samples which each quasar belongs to are
indicated by 1, 2 and 3 corresponding to VAR, UVX and AMP respectively,
as defined above.  Data for many of the quasars have already been
published by Hawkins \& V\'{e}ron (\cite{haw95}), but are given again
here for completeness.  Any small differences in the parameters are the
result of further refinement of the calibration, and more extended
monitoring of the light curves.\\

\begin{figure}
 \resizebox{\hsize}{!}{\includegraphics{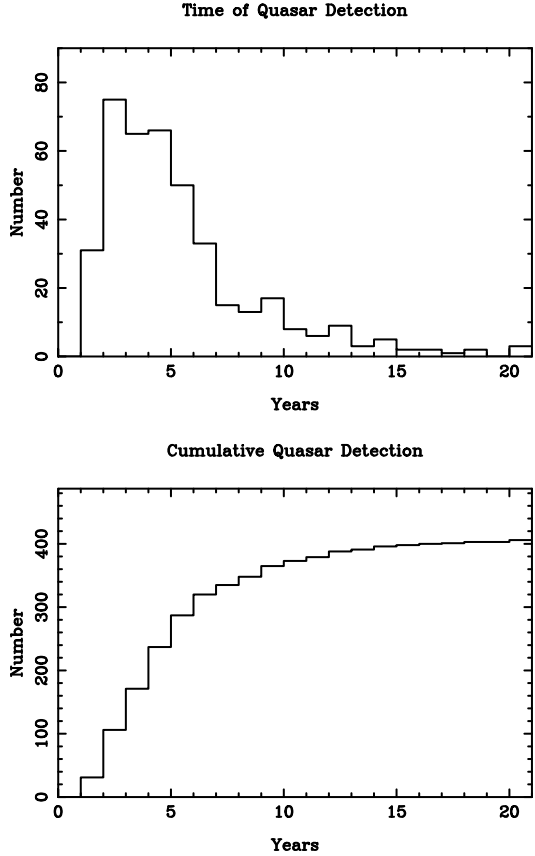}}
 \caption{The top panel shows a histogram of the epochs at which
quasars could first have been detected on the basis of the variability
criterion.  The bottom panel shows the same plot in cumulative form.}
 \label{1}
\end{figure}

UVX selected samples have been used many times in the past for quasar
surveys, and the constraints are quite well known.  Variability
selection has been less often used, and some additional comments are
appropriate.  In particular, there is the important question of
completeness.  Hawkins \& V\'{e}ron (\cite{haw95}) use a small sample of
quasars in field 287 from Morris et al. (\cite{mor91}) selected by
objective prism to test completeness.  In 1993, 79\% of the objective
prism quasars would have also been detected according to the variability
criterion $\delta m > 0.35$; by 1997 this figure had risen to 93\%,
with two quasars remaining below the variability threshold.  In fact
both of these quasars are clearly variable, with amplitudes 0.32 and
0.29 magnitudes.  Strangely, they lie at either extreme of the
redshift and luminosity range of the sample, with redshifts 3.23,
0.29 and luminosities -27.55, -22.33 respectively. 
Fig.~\ref{1}(a) shows the distribution of epoch at which quasars first
satisfy the detection criterion given by Hawkins (\cite{haw96}).  This
is generally speaking equivalent to attaining an amplitude of 0.35
magnitudes.  The distribution peaks between 2 and 4 years, and most
quasars have satisfied the criterion after 7 years.  Fig.~\ref{1}(b)
shows the cumulative distribution of detection epochs, illustrating
the point that after 15 years nearly all quasars have varied
sufficiently to put them in the variability selected sample.\\

\begin{figure}
 \resizebox{\hsize}{!}{\includegraphics{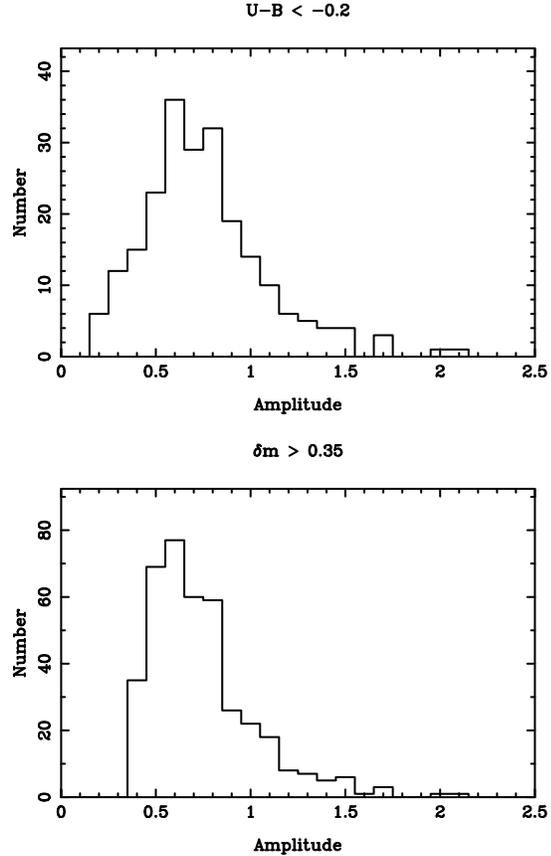}}
 \caption{The top panel shows a histogram of amplitudes for the complete
UVX detected sample.  The bottom panel shows a similar histogram for
the variability detected sample (VAR).  Quasars are constrained to
have an amplitude greater than 0.35 mags.}
 \label{2}
\end{figure}

The distribution of amplitudes is best shown with the UVX sample,
which was selected without any reference to variability.
Fig. ~\ref{2}(a)
is a histogram of amplitudes over a 20 year baseline, and shows a
median amplitude of around 0.7 magnitudes.  As would be expected
from Fig.~\ref{1}, nearly all of the sample lies above the threshold
amplitude of the variability selected sample of 0.35 mags.  With
this in mind it is worth investigating the possibility of using
the variability selected sample to look for correlations at higher
redshift.  Fig. ~\ref{2}(b) shows a histogram of amplitudes for the VAR
sample, which from the way it was selected has a cut-off at an
amplitude of 0.35 mags.  More interestingly, the distribution as
a whole peaks at an amplitude of about 0.7, similar to the UVX
sample.  This peak is well clear of the cut-off, and implies
that only a small fraction of quasars are missed from the VAR sample.
There may however be a problem detecting the most luminous quasars,
($M_{B} < -27$) for which there is evidence for relatively small
variations (Cristiani et al. \cite{cri96}).  Nonetheless, the sample
can be used with caution to extend quasar correlations to high
redshift.\\

\section{Amplitude correlations}

\begin{figure*}
 \resizebox{\hsize}{!}{\includegraphics{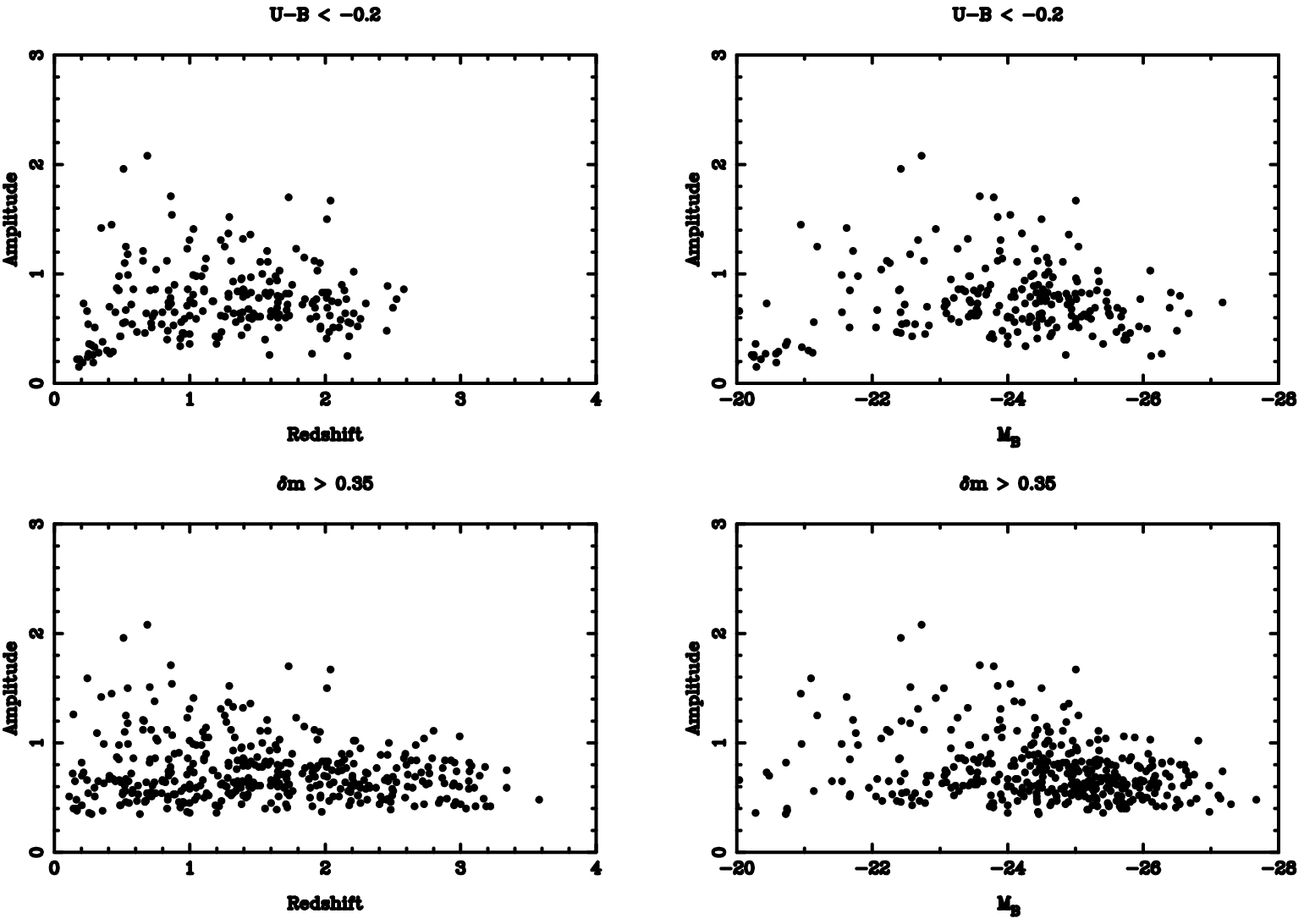}}
 \caption{The top two panels show plots of amplitude versus redshft
and absolute magnitude for quasars selected solely on the basis
of ultra-violet excess.  The bottom two panels are similar plots
for quasars selected on the basis of variability.}
 \label{3}
\end{figure*}

\begin{figure*}
 \resizebox{\hsize}{!}{\includegraphics{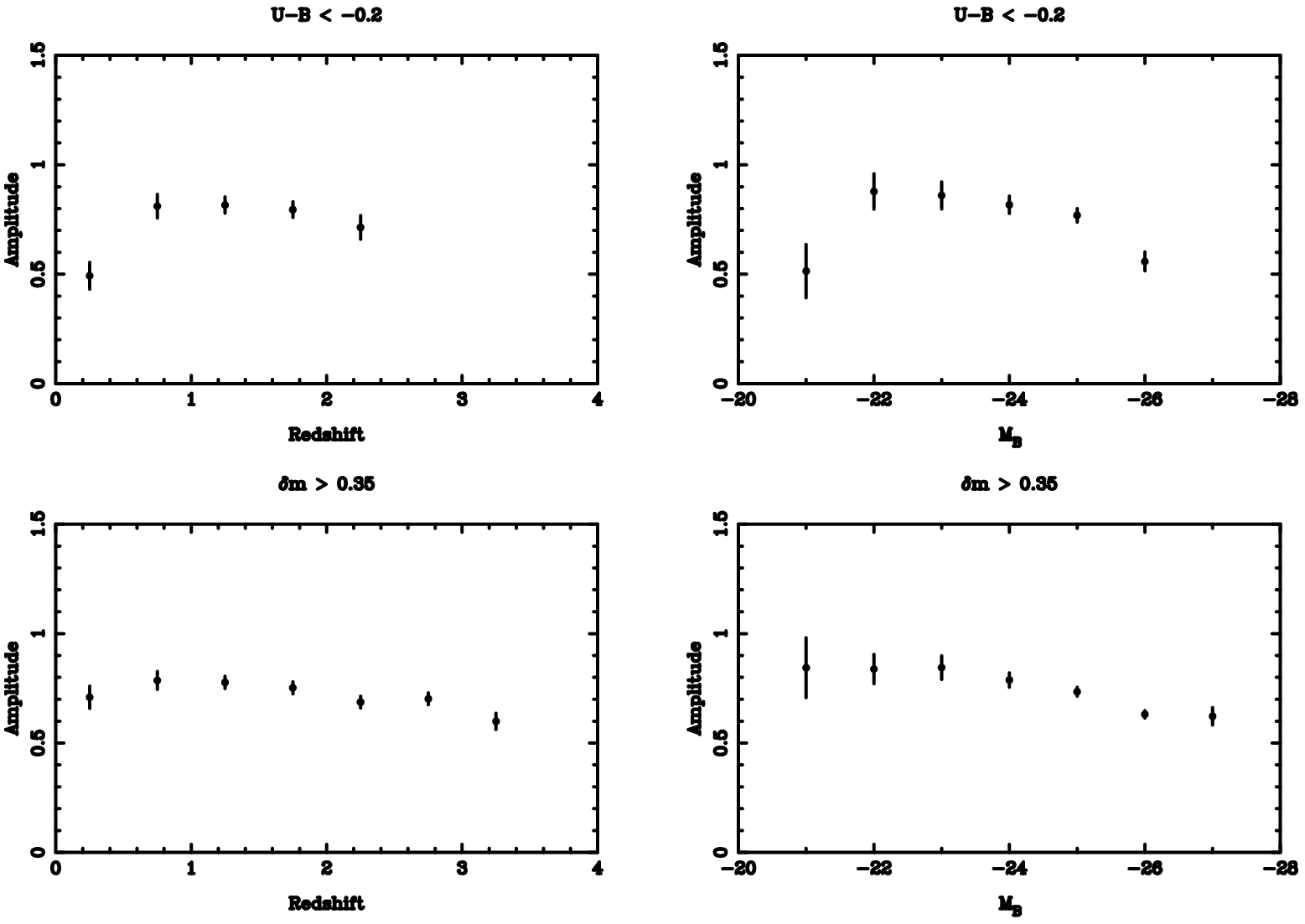}}
 \caption{Plots based on the same data as for Fig. 3, but binned in
redshift intervals of 0.5 and unit absolute magnitude intervals.
The error bars show the uncertainty in the position of the mean.}
 \label{4}
\end{figure*}

The top two panels in Fig. ~\ref{3} show the distribution of amplitudes
as a function of redshift and luminosity for the UVX sample.  The
most striking thing about these figures is the marked drop in amplitude
towards low redshift and luminosity.  This is illustrated more clearly
in the top two panels of Fig. ~\ref{4}, where the data is binned in
intervals
of 0.5 in redshift and unit absolute magnitude, and the mean plotted
with $\sqrt{N}$ error bars.  The fall towards low redshift and
luminosity is highly significanti (a 3-$\sigma$ effect), although it is
not clear from these data alone whether it is primarily a luminosity or
redshift effect.  The correlation coeffecient for the top left panel of
Fig. ~\ref{3} with $z < 0.5$ is 0.46, and for the top right panel with
$M_{B} > -22$ is -0.67, confirming the reality of the correlation.
There is also some evidence for a decrease in amplitude for the most
luminous quasars ($M_{B} < -25$), and an even weaker decline for high
redshift objects.  Correlation coefficients for all the data in the
top two panels are 0.12 and -0.10 for left and right panels
respectively, emphasising the weakness of the effect.   This is another
manifestation of the well known degeneracy between redshift and
luminosity.  This point will be investigated further below by dividing
the data into sub-samples.\\

The bottom two panels in Figs ~\ref{3} and ~\ref{4} show similar plots
for the VAR sample.  The greater redshift range allowed by variability
selection still shows little trend of amplitude with
redshift, but the decrease in amplitude for luminous quasars is
shown to continue to greater luminosities, although the effect is
inevitably lessened by the absence of quasars with
$\delta m < 0.35$.\\

\begin{figure*}
 \resizebox{\hsize}{!}{\includegraphics{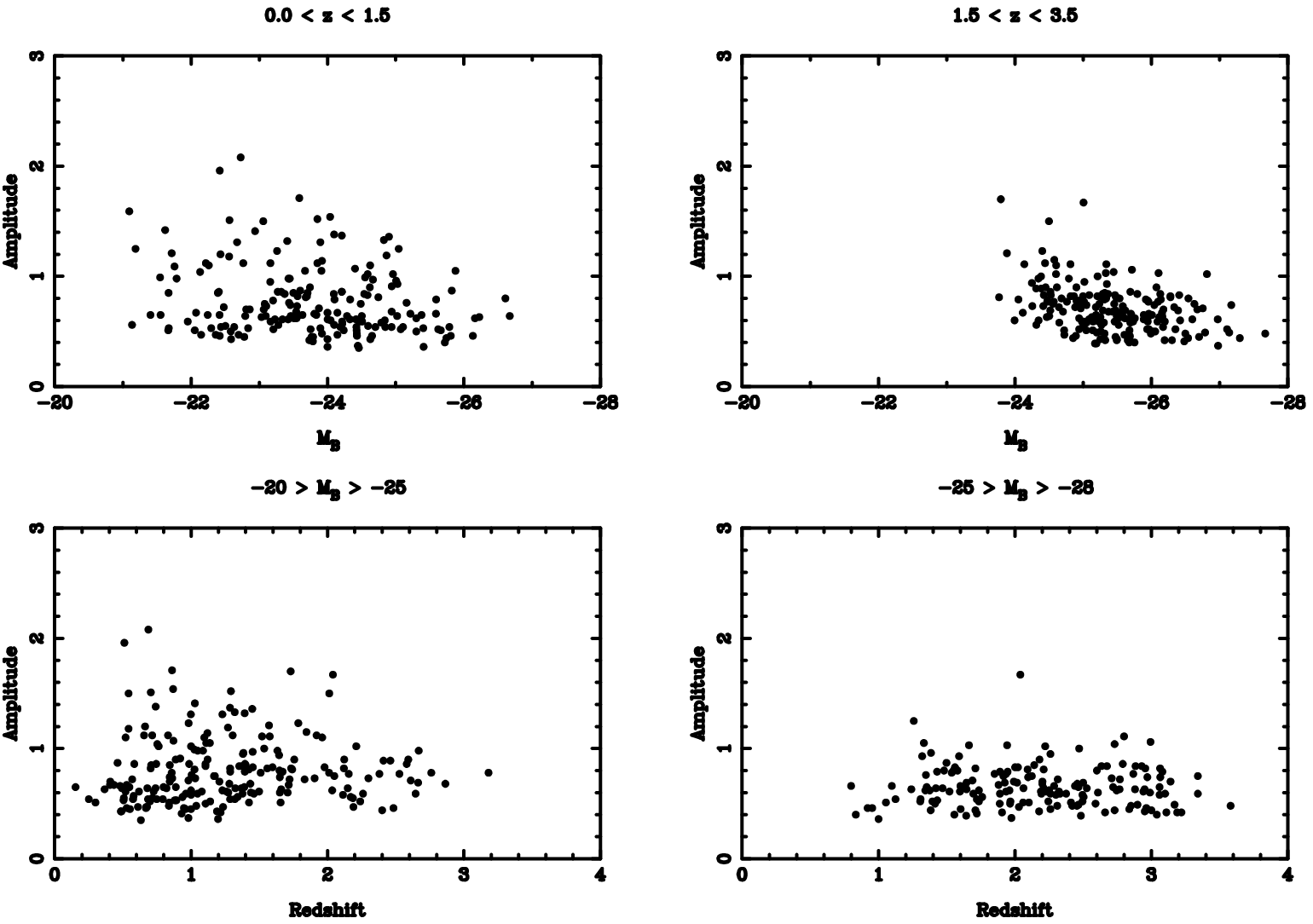}}
 \caption{Plots of amplitude versus redshift and absolute magnitude
for the variability detected sample (VAR).  The top two panels
show amplitude versus absolute magnitude for two redshift ranges,
and the bottom two panels show amplitude versus redshift for
two absolute magnitude ranges.}
 \label{5}
\end{figure*}

\begin{figure*}
 \resizebox{\hsize}{!}{\includegraphics{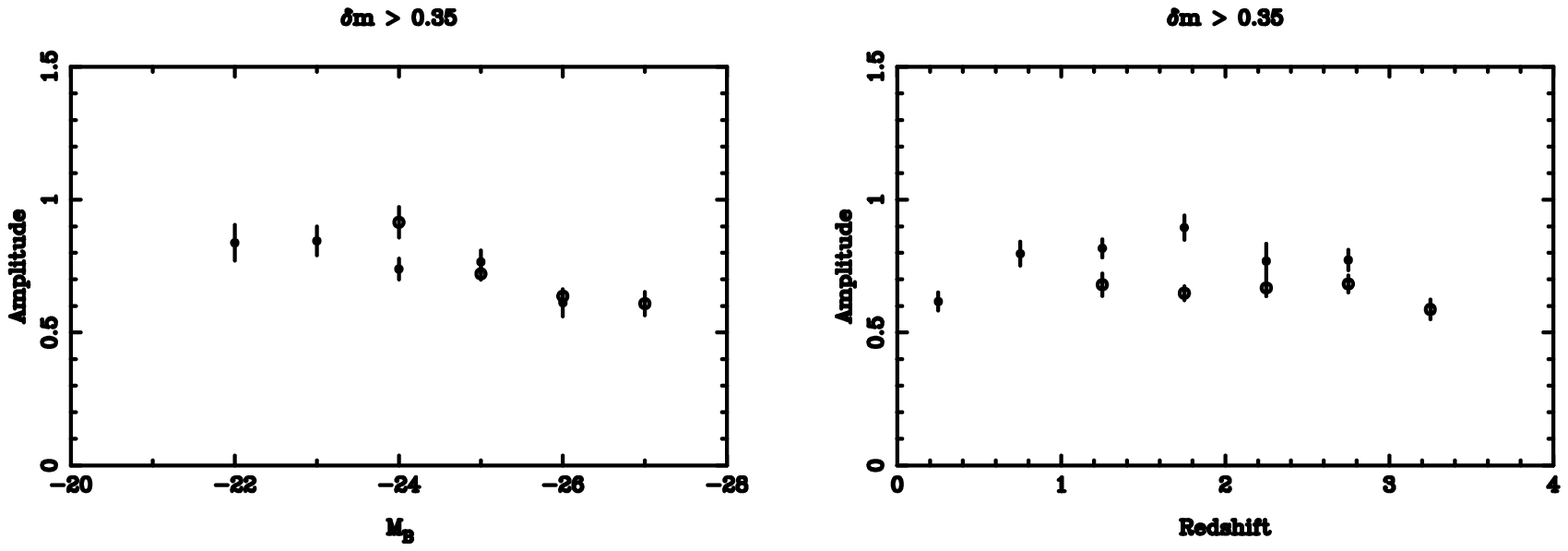}}
 \caption{Plots based on the same data as for Fig. 5, but binned in
unit absolute magnitude intervals and redshift intervals of 0.5.
The left hand panel shows the relation between amplitude and
absolute magnitude for quasars with $z < 1.5$ (small dots)
and $z > 1.5$ (large dots).  The right hand panel shows the
relation between amplitude and redshift for quasars with
$M_{B} > -25$ (small dots) and $M_{B} < -25$ (large dots).
The error bars show the uncertainty in the position of the mean.}
 \label{6}
\end{figure*}

It has been mentioned above that there is a degeneracy between
redshift and luminosity.  This results from the fact that in a
magnitude limited sample high redshift quasars tend to be high
luminosity objects, and vice versa.  Thus a trend with one parameter
will be mimicked by a trend with the other, and the true relation will
be hard to disentangle.  This degeneracy can in principle be broken byi
binning the data in redshift and luminosity, and the result of doing
this is shown in Figs. ~\ref{5} and ~\ref{6}.  The VAR sample was used
as it covers a wider range of luminosity and redshift, making binning
feasible.  The left hand panel shows amplitude as a function of
luminosity in two redshift bins, $z < 1.5$ and $z > 1.5$.  Data for
the two redshift ranges overlap nicely, and clearly show a decrease
in the relation between amplitude and luminosity.  The right hand
panel shows amplitude as a function of redshift in two luminosity
bins.  In this case there is some evidence for an increase in the
relation at low redshift, but it is essentially flat beyond $z = 0.5$
for both luminosity ranges.  It is however worth noting that the less
luminous quasars have larger amplitudes, as expected from the data in
the left hand panel.  It would thus appear that while for $M_{B} > -25$
amplitude does indeed decrease with luminosity, it does not change
with redshift for $z > 0.5$.\\

\begin{figure*}
 \resizebox{\hsize}{!}{\includegraphics{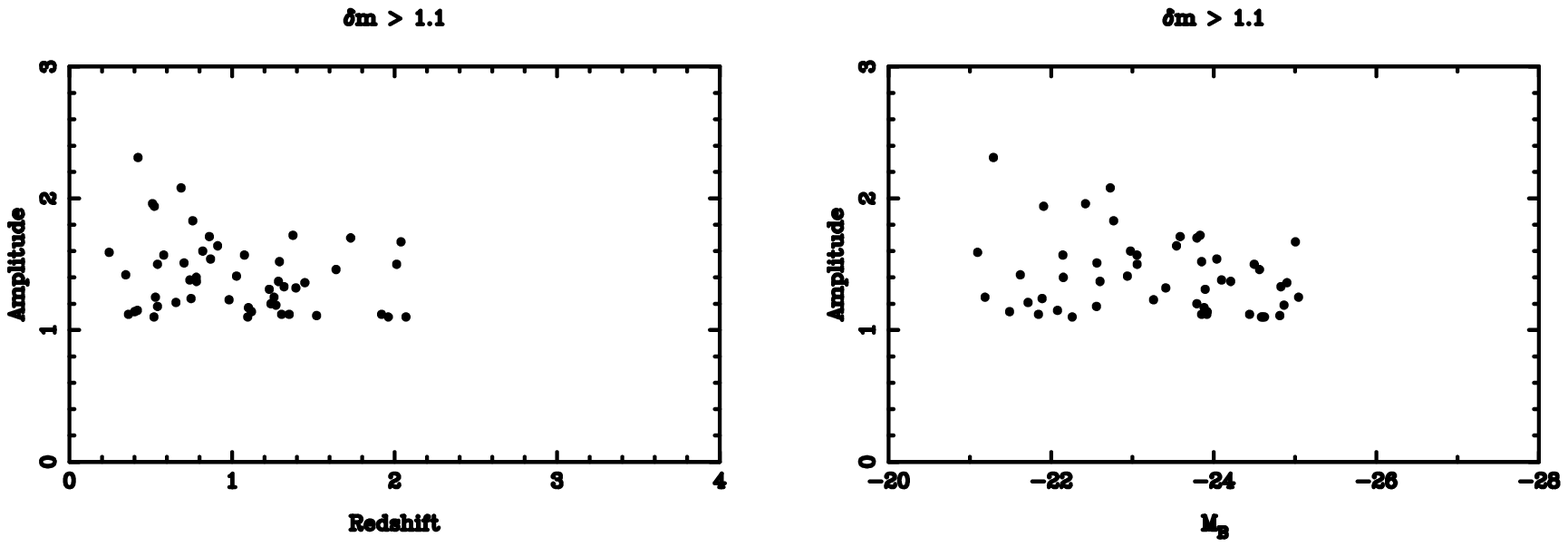}}
 \caption{Plots of amplitude versus redshift and absolute magnitude
for the most variable quasars in the sample.}
 \label{7}
\end{figure*}

Examination of Fig. ~\ref{3}, especially the bottom panels, suggests
that there may be population of low luminosity and/or low redshift
quasars distinct from the parent population.  This is particularly
evident in the bottom left hand panel of Fig. ~\ref{3}, which is
uniformly populated between amplitudes of 1.1 and 1.8 up to a redshift
of 2 at which point
there is a sharp cut-off with no amplitudes greater than 1.1 at higher
redshift.  To investigate this population with better statistics all
variables with $\delta m > 1.1$ in the measured area of field 287 were
observed on the 3.6m at La Silla to confirm their identification as
quasars and measure redshifts.  This became the AMP sample.
Fig. ~\ref{7}
shows amplitude as a function of both redshift and luminosity, and it
will be seen that there is indeed a cut-off in redshift at $z \sim 2$
and $M_{B} \sim -25$.  It is clear that this cut-off must in fact
be related to luminosity.  If it were a redshift cut-off there is
no reason why such objects should not be seen with greater luminosity.
On the other hand if it were a luminosity cut-off, then this
combined with a magnitude limit will indeed produce an effective
cut-off in redshift.\\

\section{Discussion}

\begin{figure}
 \resizebox{\hsize}{!}{\includegraphics{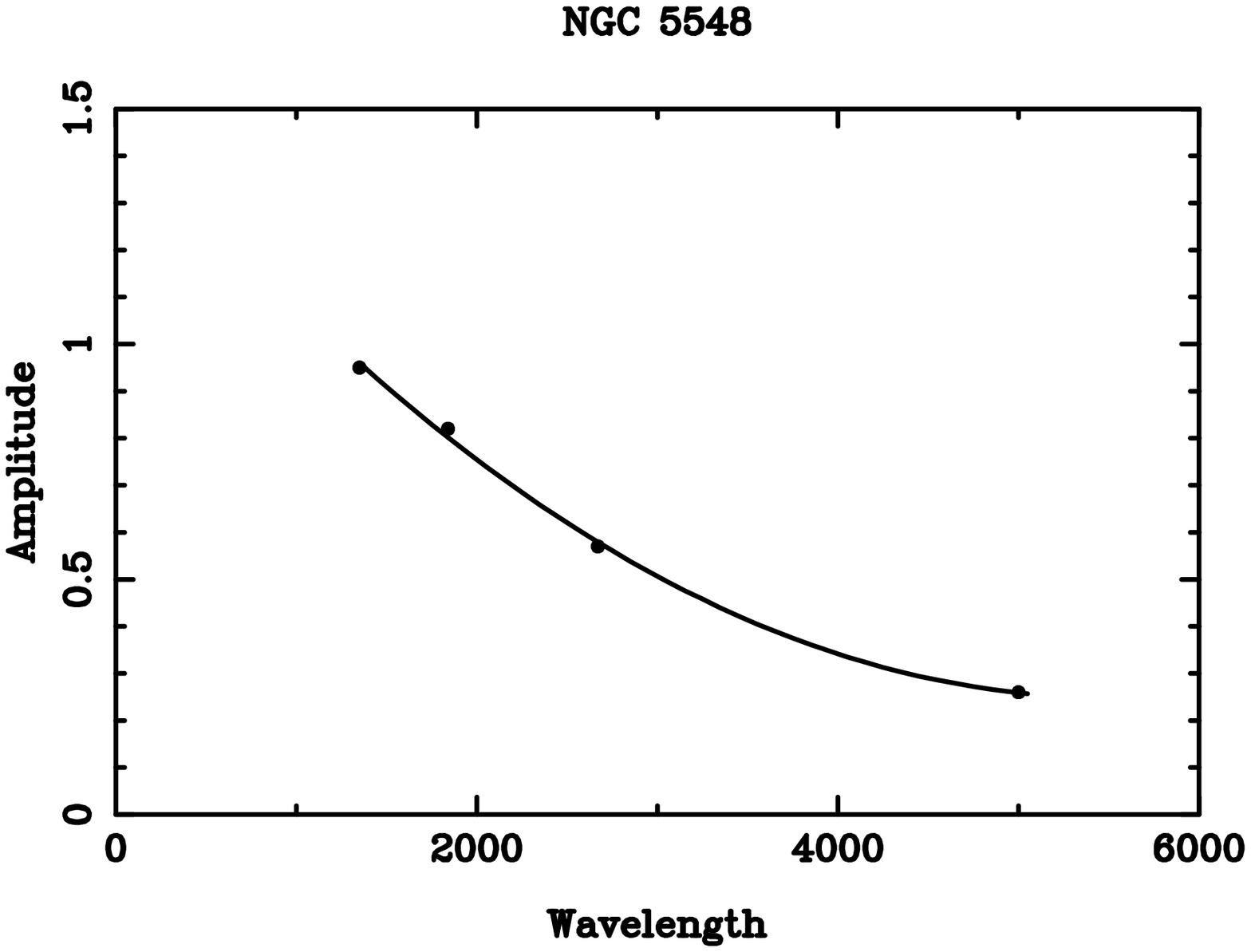}}
 \caption{Relation between amplitude and wavelength for the Seyfert
galaxy NGC 5548 from IUE data published by Clavel et al. 1991.}
 \label{8}
\end{figure}

Attempts to measure correlations of amplitude (or some related 
variability parameter) with redshift and luminosity have a long history,
which is summarised by Hawkins (\cite{haw96}).  Among recent work, a
useful place to start is with the paper by Hook et al. (\cite{hoo94}).
They analyse a sample of $\sim 300$ quasars in the SGP area from 12 UK
1.2m Schmidt plates taken in 5 separate yearly epochs spanning 16 years.
They find a convincing anti-correlation between their variability
parameter $\sigma_{v}$ (a measure of variation about the mean) and
luminosity, but a much weaker anti-correlation with redshift.  They
attribute this to the degeneracy between redshift and luminosity
in their sample.  The same sample is re-analysed by Cid Fernandes
et al. (\cite{cid96}) who use a variability index related to variance,
and also one related to the structure function.  They concur with
Hook et al. (\cite{hoo94}) that there is an anti-correlation between
their variability indices and luminosity, but claim a positive
correlation with redshift.  This effect is not apparent to the
eye, but is interpreted as a variability-wavelength dependence
rather than an intrinsic variability-redshift dependence.  The
net result in an un-binned sample cancels out with the luminosity
variability relationship.  Cid Fernandes et al.'s claim for a
positive correlation between amplitude and redshift appears to be
motivated at least in part by expectations arising from a paper by
Di Clemente et al. (\cite{dic96}).  This interesting paper examines
the relation between their variability parameter $S_{1}$ (an amplitude
based on the structure function) and wavelength.  Their sample is
composed of PG quasars (Schmidt and Green \cite{sch83}), which are
mostly low redshift and relatively low luminosity objects.  With the
help of archival IUE observations they find that $S_{1}$ decreases
with wavelength.  This effect can clearly be seen in the light curve
from the intensive monitoring programme of the Seyfert galaxy NGC 5548
(Clavel et al. \cite{cla91}), which has a larger amplitude at shorter
wavelength.  Fig. ~\ref{8} shows the relation between amplitude and
wavelength taken from the light curves, with a best-fit quadratic
curve.\\

\begin{figure}
 \resizebox{\hsize}{!}{\includegraphics{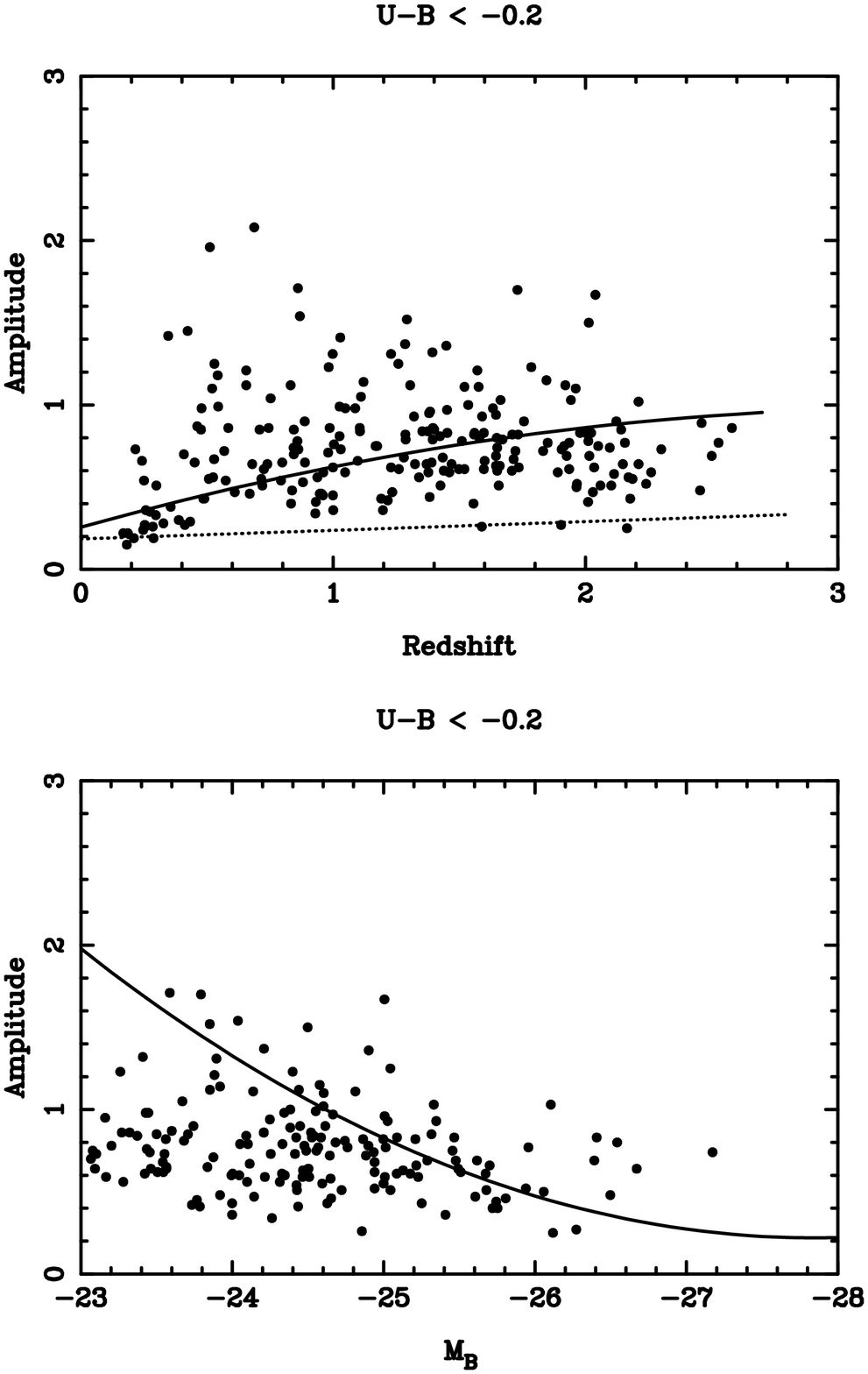}}
 \caption{Plots of amplitude versus redshift and luminosity for UVX
selected quasars.  In the top panel the solid line represents
the relation from Fig. 8, with wavelength converted to redshift;
the dotted line is the model from Di Clemente et al. (1996).}
 \label{9}
\end{figure}

The relation between rest wavelength and amplitude is essentially
equivalent to the relation between redshift and amplitude, where one
is seeing progressively shorter wavelengths at higher redshift.  This
is illustrated in the top panel of Fig. ~\ref{9} which shows the UVX
sample with two curves superimposed.  The solid line is converted from
Fig. ~\ref{8} and does not appear to follow the trend of the data,
but the large scatter makes it hard to construct a convincing test.
Nonetheless it suggests that any relation which holds for Seyfert
galaxies might have to be modified for quasars.  The dotted line is
from Di Clemente et al. (\cite{dic96}), and shows a very small effect,
which does nonetheless follow the flat distribution of the data.\\

The decrease of amplitude for the smallest redshift and luminosity
seen in the top two panels of Fig. ~\ref{4} has a number of possible
explanations.  It could be the effect of the underlying galaxy
dominating any change in nuclear brightness for low luminosity
objects, it could be a consequence of the small optical depth to
microlensing at low redshift, or it could be a consequence of the
wavelength dependence of variability (Cristiani et al. \cite{cri97}).
The present dataset is not adequate to settle the question, which
is best done by looking at luminous quasars at very low redshift
and Seyfert galaxies at high redshift.\\

All the plots in Figs ~\ref{3} and ~\ref{4} show a trend of decreasing
amplitude
towards higher redshift or more luminous objects.  Although
the trend as a function of luminosity is more marked, the old
problem of degeneracy makes it hard to say for certain that it is
a luminosity effect which is being observed.  However, if we look
at Fig. ~\ref{6} where the data are binned in luminosity and redshift
we see that while there is no significant trend of amplitude with
redshift in either luminosity bin, there is a marked inverse
correlation between amplitude and luminosity.\\

The relation between amplitude and luminosity is in agreement with that
found in earlier work (Hook et al. \cite{hoo94}, Hawkins \cite{haw96},
Cristiani et al. \cite{cri96}) and may well turn out to be a useful way
of distinguishing between various schemes for quasar variability.
The evidence for a constant amplitude with redshift is more
debatable.  It would appear to be consistent with the early
claim of Hook et al. (\cite{hoo94}) for a weak anti-correlation with
redshift which they ascribed to degeneracy with luminosity.  It
is also in agreement with the results of Cristiani et al. in
the observer's frame.  When they correct their structure function
to the quasar rest frame they inevitably imprint a positive
correlation between their variability parameter and redshift.\\

\begin{figure}
 \resizebox{\hsize}{!}{\includegraphics{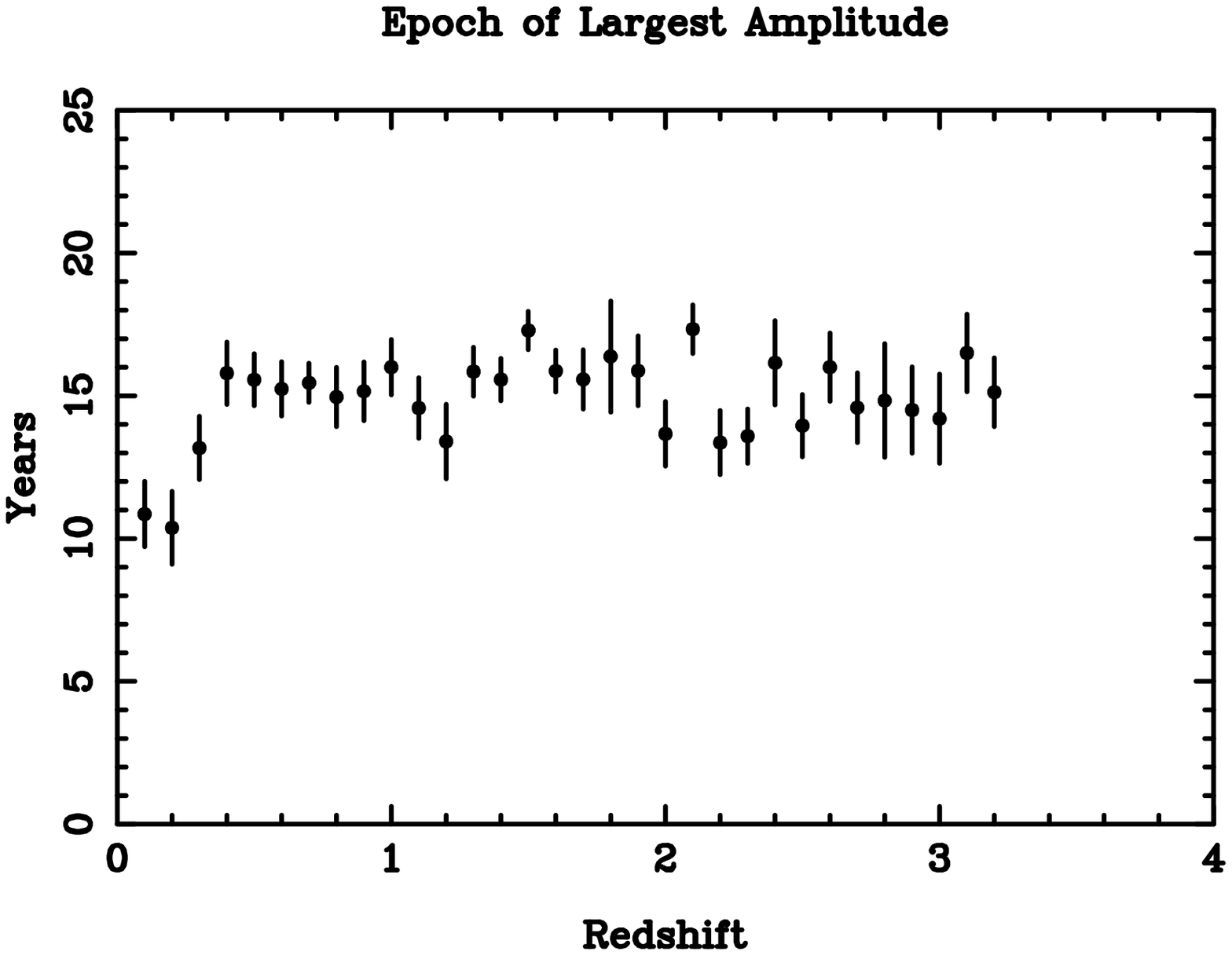}}
 \caption{Plot of the epoch at which the largest magnitude difference
is achieved in quasar light curves at different redshifts.  Poisson
error bars are shown based on the number of objects in each bin.}
 \label{10}
\end{figure}

To investigate this dependence of amplitude on redshift for the
present sample, Fig. ~\ref{10} shows the epoch at which quasars achieve
their maximum amplitude as a function of redshift.  Apart from
very low redshifts ($z < 0.3$) this relation is flat, implying
that at least over the 21 years of the present dataset, time
dilation effects will not bias the measurement of amplitude.  Since
the conclusions of Cid Fernandes et al. (\cite{cid96}) are largely
based on a sample for which a time dilation correction has been
applied, it is not feasible to make a direct comparison with the 
present work.  However, it appears that the main difference
between their results and those of Hook et al. is in the definition
of a variability parameter and the method of analysis (both papers
are based on the SGP sample).\\

There are perhaps three currently discussed schemes for quasar
variability.  The least well constrained is the accretion disk model,
where instabilities are propagated across the disk leading to
variation in light.  The details of this approach have proved hard to
work out, especially in the context of the constraints imposed by
existing observations, but it does not seem to lead to an inverse
correlation between amplitude and luminosity.  An interesting recent
attempt to model variation on the basis of accretion disk
instabilities by Kawaguchi et al. (\cite{kaw98}) may provide a means for
producing the observed variations.  It does however appear to predict
variations which are either too asymmetric or of too small an
amplitude to be consistent with the current observations.  The
timescales which they predict are also rather short, around 200
days for reasonable input parameters, and much shorter than the
observed timescale of a few years.\\

An alternative approach, developed by Terlevich and his collaborators,
accounts for the variation by postulating that the quasar is powered
by a series of supernova explosions.  Qualitatively, this model can
account for the observed relation between luminosity and amplitude,
and works quite well for Seyfert galaxies (Aretxaga \& Terlevich
\cite{are94}).  However, for quasars (Aretxaga et al. \cite{are97})
large numbers of supernovae are required to achieve the luminosity,
which results in smaller variations.  For example, even a relatively
modest quasar with absolute magnitude $M_{B} \sim -26$ would need
some 300 type II supernovae per year to power it, which given typical
decay times would lead to very little variation at all.  This clearly
conflicts with observations in this paper which show that most quasars
vary by around 0.5 to 1 magnitude on a timescale of a few years.\\

The third way of explaining quasar variability is to invoke
microlensing.  This approach has been explored in several recent papers
(Hawkins \cite{haw93}, \cite{haw96}, Hawkins \& Taylor \cite{haw97}),
and seems to account well for a number of statistical properties of
quasar light curves.  It can also explain the inverse correlationi
between luminosity and amplitude in a natural way.  It is well known
that when a point source is microlensed by a population of compact
bodies of significant optical depth, the lenses combine non-linearly
to form a caustic pattern which produces sharp spikes in the resulting
light curves (Schneider \& Weiss \cite{sch87}).  As the source becomes
comparable in size with, or larger than, the Einstein radius of the
lenses the amplitude decreases (Refsdal \& Stabell \cite{ref91}).
Thus if one assumes a uniform temperature for quasar disks, and that
the luminosity is determined by the disk area, the larger more luminous
disks will be amplified less.  Using a relation between source size
(in terms of Einstein radius) and amplitude given by Refsdal \& Stabell
(\cite{ref91}) (eqn 1), one can thus derive a relation between
amplitude $\delta m$ and absolute magnitude $M$ of the form:

\[ \delta m = 10^{0.2(M+c)} \]

\noindent
where $c$ is a constant.  The bottom panel of Fig. ~\ref{9} shows a
plot of amplitude versus absolute magnitude for the UVX sample with this
relation superimposed.  The constant $c$ was adjusted to allow the
curve to track the upper envelope of the points, which has the effect
of defining the quasar disk size.  This ranges from 1.1 Einstein radii
for $M_{B} = -23$ to 11 for $M_{B} = -28$.  The points scatter
downwards from the upper envelope because the quasars do not
necessarily attain their maximum possible amplitude.  Although one
cannot say that the curve provides a fit to the data, the trend is
certainly well represented.\\

The possibility of a distinct population of large amplitude, low
luminosity quasars suggested by Fig. ~\ref{7} may also be used to test
the models of quasar variability.  Again, an accretion disk provides
no obvious mechanism for such an effect.  The Christmas Tree model
certainly does imply large amplitude variations for low luminosity
objects, where each event is of comparable brightness to the nucleus
itself.  Perhaps the most natural explanation comes from microlensing,
where the nucleus of low luminosity quasars would plausibly become
very small compared with the Einstein radii of the lenses, resulting
in large amplifications from caustic crossing events.\\

\section{Conclusions}

In this paper we have examined the dependance of amplitude on redshift
and luminosity for large samples of quasars selected on the basis
of ultra-violet excess and variability.  The quasars span a redshift
range $0 < z < 3.5$ and a luminosity range $-20 > M_{B} > -28$.
There is evidence for a correlation of amplitude with luminosity and/or
redshift for the sample as a whole, but when it is binned in redshift
the correlation with luminosity becomes significant.  This result could
be strengthened by the possible existence of a population of large
amplitude low luminosity objects in the sample.  No convincing evidence
is found for a correlation between amplitude and redshift, either for
the sample as a whole or when it is binned in luminosity.\\

Various models of quasar variability are examined with respect
to the observed correlations.  It is concluded that any straightforward
interpretation of an accretion disk model is incompatible with the
data.  The Christmas Tree model has some merits, especially in the
regime of low luminosity quasars and Seyfert galaxies, but for
luminous quasars the rate of supernovae required is too large to
be compatible with the observed variability.  The microlensing model
can be used to explain all the data, although it does require that
the Einstein radius of the microlensing bodies is comparable in size
to the quasar nucleus.\\

\begin{acknowledgements}

I thank Andy Lawrence and Omar Almaini for making some excellent
suggestions for improvements to the paper.

\end{acknowledgements}

\pagestyle{empty}

\section*{Appendix}

\begin{table}[h]

\begin{tabular}[b]{rrrrrrrrrrrrrr}
\multicolumn{3}{c}{R.A. (1950)}&\multicolumn{3}{c}{Dec (1950)}&
\multicolumn{1}{c}{$B$}&\multicolumn{1}{c}{$U\!-\!B$}&
\multicolumn{1}{c}{$\delta m$}& \multicolumn{1}{c}{$z$}&
\multicolumn{1}{c}{$M_{B}$}&\multicolumn{3}{c}{Samples}\\
 &&&&&&&&&&&&&\\
 21& 16& 28.38&  -44&  0& 57.3& 19.34& -0.42&  0.64& 2.062& -26.04& 1& 0& 0\\
 21& 16& 22.50&  -44& 29& 13.9& 19.33& -0.77&  1.19& 1.270& -25.03& 1& 0& 3\\
 21& 16& 44.32&  -44&  9& 28.0& 20.29& -0.59&  1.72& 1.375& -24.24& 0& 0& 3\\
 21& 16& 13.41&  -46& 42& 47.3& 18.71& -0.40&  0.87& 1.498& -26.00& 1& 0& 0\\
 21& 16& 25.78&  -46& 10& 42.1& 18.92& -0.37&  0.54& 0.748& -24.32& 1& 0& 0\\
 21& 16& 55.05&  -44& 39& 37.1& 17.89& -0.09&  0.80& 1.480& -26.80& 1& 0& 0\\
 21& 17& 24.47&  -42& 41&  4.1& 19.51& -0.35&  1.12& 0.363& -22.17& 0& 0& 3\\
 21& 16& 46.14&  -46& 14& 21.7& 19.08& -0.23&  0.59& 0.297& -22.17& 1& 0& 0\\
 21& 17& 30.01&  -44&  2& 40.4& 18.30& -0.45&  0.44& 1.710& -26.69& 1& 0& 0\\
 21& 17& 25.61&  -45&  4& 58.1& 20.40&  0.25&  0.72& 2.313& -25.22& 1& 0& 0\\
 21& 17& 26.97&  -45& 30& 20.9& 20.28&  1.15&  1.06& 2.993& -25.86& 1& 0& 0\\
 21& 17& 21.81&  -47&  3& 49.0& 18.80&  0.01&  0.45& 2.260& -26.77& 1& 0& 0\\
 21& 17& 53.12&  -46&  0& 44.8& 19.71&  0.31&  0.82& 2.955& -26.41& 1& 0& 0\\
 21& 17& 55.61&  -46&  9& 17.1& 20.38&  0.10&  0.82& 2.120& -25.06& 1& 0& 0\\
 21& 17& 48.23&  -46& 47& 36.2& 19.35& -0.50&  0.48& 1.038& -24.59& 1& 0& 0\\
 21& 17& 57.12&  -46& 20& 47.4& 20.50&  0.46&  0.84& 2.884& -25.57& 1& 0& 0\\
 21& 18& 21.07&  -44& 59& 51.9& 19.19& -0.12&  0.59& 2.194& -26.32& 1& 0& 0\\
 21& 18& 34.84&  -43& 49& 45.2& 19.29& -0.43&  0.90& 1.121& -24.81& 1& 0& 0\\
 21& 18& 34.30&  -44&  0& 44.9& 20.44&  1.01&  0.70& 3.140& -25.80& 1& 0& 0\\
 21& 18& 41.77&  -43& 24& 38.6& 20.78& -0.47&  1.57& 1.076& -23.23& 0& 0& 3\\
\end{tabular}

\end{table}

\begin{table*}[t]

\begin{tabular}[b]{rrrrrrrrrrrrrr}
\multicolumn{3}{c}{R.A. (1950)}&\multicolumn{3}{c}{Dec (1950)}&
\multicolumn{1}{c}{$B$}&\multicolumn{1}{c}{$U\!-\!B$}&
\multicolumn{1}{c}{$\delta m$}& \multicolumn{1}{c}{$z$}&
\multicolumn{1}{c}{$M_{B}$}&\multicolumn{3}{c}{Samples}\\
 &&&&&&&&&&&&&\\
 21& 18& 47.48&  -43&  1&  9.5& 18.81& -0.62&  0.75& 2.123& -26.63& 1& 0& 0\\
 21& 18& 11.92&  -46& 56&  3.6& 18.35& -0.37&  0.41& 1.720& -26.65& 1& 0& 0\\
 21& 18& 55.69&  -43&  3& 36.1& 18.66& -0.01&  0.70& 2.201& -26.85& 1& 0& 0\\
 21& 19&  0.54&  -43& 48&  0.5& 20.72&  0.63&  0.82& 3.064& -25.47& 1& 0& 0\\
 21& 18& 29.22&  -47&  2& 34.9& 18.36& -0.42&  1.05& 1.332& -26.10& 1& 0& 0\\
 21& 19& 14.89&  -43&  7& 27.0& 19.81&  0.43&  0.84& 2.929& -26.29& 1& 0& 0\\
 21& 19& 37.27&  -43& 23&  9.2& 20.31&  0.07&  1.04& 2.730& -25.65& 1& 0& 0\\
 21& 19& 20.05&  -45& 30& 52.3& 20.38&  0.00&  0.51& 2.410& -25.32& 1& 0& 0\\
 21& 19& 45.46&  -43& 32& 19.3& 20.65& -0.65&  1.40& 0.652& -22.29& 0& 0& 3\\
 21& 19& 47.61&  -43& 46& 23.9& 19.16& -0.40&  1.02& 1.000& -24.70& 1& 0& 0\\
 21& 19& 37.16&  -45& 48& 22.3& 19.98&  0.03&  0.81& 2.210& -25.54& 1& 0& 0\\
 21& 19& 55.45&  -44& 17& 54.7& 20.73&  0.59&  0.79& 3.080& -25.47& 1& 0& 0\\
 21& 20& 14.92&  -42& 55& 30.5& 20.89&  0.15&  0.53& 0.619& -21.94& 1& 0& 0\\
 21& 19& 49.62&  -45& 49&  8.4& 20.65& -0.26&  0.67& 1.715& -24.35& 1& 2& 0\\
 21& 20& 18.19&  -44& 10& 40.6& 20.69& -0.10&  1.10& 0.580& -22.00& 0& 0& 3\\
 21& 20& 28.49&  -43& 38& 36.6& 19.23& -0.35&  0.51& 0.840& -24.26& 1& 0& 0\\
 21& 20&  4.87&  -46& 34& 17.3& 19.56&  0.02&  0.59& 2.375& -26.11& 1& 0& 0\\
 21& 20& 21.60&  -45& 52& 12.2& 20.22&  0.85&  0.60& 2.989& -25.92& 1& 0& 0\\
 21& 20& 50.81&  -43& 27& 55.6& 17.92& -0.49&  0.63& 1.240& -26.39& 1& 0& 0\\
 21& 20& 58.76&  -44& 13& 48.6& 20.94&  0.02&  0.81& 1.741& -24.09& 1& 0& 0\\
 21& 20& 47.86&  -45& 32& 44.2& 20.03&  1.10&  0.61& 2.941& -26.08& 1& 0& 0\\
 21& 20& 57.48&  -46& 20& 36.2& 20.09& -0.35&  0.64& 0.739& -23.12& 1& 2& 0\\
 21& 21& 26.81&  -43& 51& 34.2& 19.20& -0.49&  0.79& 1.946& -26.06& 1& 0& 0\\
 21& 21& 20.68&  -44& 45&  7.7& 20.29&  0.98&  0.46& 2.963& -25.83& 1& 0& 0\\
 21& 21& 19.34&  -45&  5&  0.8& 21.18& -0.07&  1.24& 0.748& -22.06& 0& 0& 3\\
 21& 21& 12.02&  -46&  4&  1.4& 20.34&  0.09&  0.95& 2.260& -25.23& 1& 0& 0\\
 21& 21& 22.69&  -45& 58& 25.6& 20.74& -0.32&  0.60& 1.650& -24.17& 0& 2& 0\\
 21& 21& 41.90&  -44&  6& 57.6& 17.80& -0.46&  0.52& 1.735& -27.22& 1& 0& 0\\
 21& 21& 33.83&  -45& 15&  9.2& 20.39& -0.11&  1.83& 0.758& -22.88& 0& 0& 3\\
 21& 21& 23.86&  -46& 38&  7.0& 19.92& -0.21&  1.64& 0.912& -23.74& 0& 0& 3\\
 21& 21& 31.37&  -46& 13& 36.3& 20.51& -0.48&  0.59& 1.047& -23.45& 0& 2& 0\\
 21& 21& 43.13&  -44& 56& 36.3& 20.40&  0.38&  0.63& 2.950& -25.71& 1& 0& 0\\
 21& 21& 40.68&  -45& 51& 23.9& 18.89& -0.43&  0.46& 0.947& -24.85& 1& 2& 0\\
 21& 21& 42.39&  -45& 49& 26.6& 20.10& -0.28&  1.10& 0.520& -22.36& 1& 2& 3\\
 21& 21& 35.46&  -46& 42& 27.4& 19.11& -0.41&  0.76& 1.347& -25.38& 1& 0& 0\\
 21& 21& 41.39&  -45& 59& 53.9& 19.73& -0.58&  0.90& 0.887& -23.87& 1& 2& 0\\
 21& 21& 39.65&  -46& 41& 17.2& 20.35& -0.52&  1.12& 1.352& -24.15& 0& 0& 3\\
 21& 21& 56.68&  -45&  8& 33.6& 18.83& -0.31&  0.65& 1.353& -25.67& 1& 0& 0\\
 21& 22& 12.50&  -43&  9& 36.5& 20.83&  0.46&  0.65& 3.060& -25.36& 1& 0& 0\\
 21& 22&  0.52&  -45& 57& 24.8& 17.75& -0.42&  0.46& 0.953& -26.01& 0& 2& 0\\
 21& 22& 25.85&  -42& 43& 24.2& 20.16&  0.17&  0.68& 2.457& -25.58& 1& 0& 0\\
 21& 22& 30.46&  -43& 28& 38.6& 20.01& -0.10&  1.14& 0.401& -21.89& 0& 0& 3\\
 21& 22& 25.60&  -44& 22& 18.5& 18.87&  0.22&  0.49& 2.465& -26.88& 1& 0& 0\\
 21& 22& 16.68&  -46& 31& 54.8& 20.27& -0.10&  1.60& 0.820& -23.17& 0& 0& 3\\
 21& 22& 39.06&  -44& 33&  4.8& 20.32&  0.05&  0.80& 2.600& -25.54& 1& 0& 0\\
 21& 22& 41.46&  -44& 47& 46.1& 19.92&  0.06&  1.05& 1.137& -24.21& 1& 0& 0\\
 21& 22& 52.08&  -43& 33& 58.6& 18.71& -0.07&  0.45& 0.552& -23.88& 1& 0& 0\\
 21& 22& 35.59&  -46& 22&  5.7& 19.06& -0.37&  0.81& 2.093& -26.35& 1& 0& 0\\
 21& 22& 48.18&  -45& 55& 30.6& 19.96& -0.53&  0.51& 1.425& -24.65& 1& 2& 0\\
 21& 22& 49.75&  -45& 58& 24.4& 19.41& -0.31&  0.86& 0.986& -24.42& 1& 2& 0\\
 21& 22& 56.31&  -45& 57& 17.2& 20.13& -0.28&  0.78& 0.858& -23.40& 1& 2& 0\\
 21& 23&  1.03&  -46&  3& 38.1& 19.42& -0.61&  0.96& 1.384& -25.13& 1& 2& 0\\
 21& 23&  3.18&  -46&  5& 48.0& 20.53&  0.70&  0.42& 3.220& -25.76& 1& 0& 0\\
 21& 23& 21.19&  -43& 38& 29.5& 18.37& -0.18&  0.66& 0.480& -23.92& 1& 0& 0\\
 21& 23&  3.71&  -46& 54& 52.7& 18.63& -0.24&  0.79& 1.429& -25.98& 1& 0& 0\\
\end{tabular}

\end{table*}

\begin{table*}[t]

\begin{tabular}[b]{rrrrrrrrrrrrrr}
\multicolumn{3}{c}{R.A. (1950)}&\multicolumn{3}{c}{Dec (1950)}&
\multicolumn{1}{c}{$B$}&\multicolumn{1}{c}{$U\!-\!B$}&
\multicolumn{1}{c}{$\delta m$}& \multicolumn{1}{c}{$z$}&
\multicolumn{1}{c}{$M_{B}$}&\multicolumn{3}{c}{Samples}\\
 &&&&&&&&&&&&&\\
 21& 23& 14.08&  -45& 48& 53.9& 19.12& -0.22&  0.80& 1.549& -25.66& 1& 0& 0\\
 21& 23& 31.47&  -42& 56& 35.7& 19.49&  0.23&  1.26& 0.141& -20.15& 1& 0& 3\\
 21& 23& 13.92&  -46& 54& 33.0& 20.34& -0.10&  2.31& 0.422& -21.67& 0& 0& 3\\
 21& 23& 37.93&  -43& 46& 58.0& 18.91& -0.44&  0.58& 1.000& -24.95& 1& 0& 0\\
 21& 23& 43.22&  -44& 29&  5.9& 19.66& -0.66&  0.84& 1.353& -24.84& 1& 2& 0\\
 21& 23& 45.16&  -44& 29&  9.8& 20.06& -0.08&  1.17& 1.586& -24.77& 0& 0& 3\\
 21& 23& 48.33&  -45& 38&  4.0& 19.50& -0.28&  0.39& 1.642& -25.40& 1& 0& 0\\
 21& 24&  7.74&  -44& 12& 53.9& 19.39& -0.16&  1.50& 0.542& -23.16& 1& 0& 3\\
 21& 24&  5.49&  -44& 43& 57.5& 19.83& -0.32&  0.26& 1.589& -25.01& 0& 2& 0\\
 21& 24&  9.52&  -44& 12& 58.3& 20.26& -0.43&  1.20& 1.239& -24.05& 0& 0& 3\\
 21& 24&  3.30&  -46& 27& 37.3& 20.78&  0.00&  0.52& 1.140& -23.36& 1& 0& 0\\
 21& 24&  7.41&  -46&  5& 44.6& 20.45&  0.06&  1.51& 0.705& -22.66& 1& 0& 3\\
 21& 24& 11.89&  -45& 40& 17.0& 18.78& -0.52&  0.52& 1.395& -25.78& 1& 0& 0\\
 21& 24& 19.46&  -45& 15&  1.7& 20.03& -0.51&  0.61& 1.935& -25.22& 1& 2& 0\\
 21& 24& 30.39&  -43&  3& 34.7& 19.46& -0.39&  0.54& 1.286& -24.93& 1& 0& 0\\
 21& 24& 33.74&  -44& 12&  3.5& 20.01& -0.34&  1.88& 1.625& -24.87& 0& 0& 3\\
 21& 24& 21.85&  -46& 52&  0.4& 19.90&  0.58&  0.58& 2.495& -25.87& 1& 0& 0\\
 21& 24& 49.75&  -43& 20& 36.0& 19.01&  0.07&  0.64& 0.820& -24.43& 1& 0& 0\\
 21& 24& 46.25&  -44& 32& 24.7& 20.66&  0.16&  0.47& 0.677& -22.37& 1& 0& 0\\
 21& 24& 46.53&  -45& 53& 46.9& 20.48& -0.62&  1.10& 1.961& -24.79& 0& 2& 3\\
 21& 24& 54.69&  -45&  4& 54.4& 20.48& -0.32&  0.30& 0.388& -21.35& 0& 2& 0\\
 21& 24& 56.51&  -45& 53&  2.6& 19.76&  0.03&  0.61& 2.322& -25.86& 1& 0& 0\\
 21& 25&  5.28&  -44& 48& 42.3& 19.91& -0.40&  0.82& 1.733& -25.11& 0& 2& 0\\
 21& 25& 25.85&  -43& 35& 13.5& 20.32&  0.01&  0.59& 2.260& -25.25& 1& 0& 0\\
 21& 25& 23.50&  -44& 24& 24.2& 20.01& -0.16&  0.69& 2.500& -25.77& 1& 0& 0\\
 21& 25& 23.30&  -45& 16& 17.9& 20.62& -0.28&  0.66& 0.242& -20.19& 1& 2& 0\\
 21& 25& 28.96&  -45&  9& 50.0& 19.41& -0.38&  0.93& 1.590& -25.43& 1& 2& 0\\
 21& 25& 28.12&  -45& 36& 53.9& 20.44&  0.24&  0.63& 2.766& -25.54& 1& 0& 0\\
 21& 25& 28.84&  -46& 39& 17.1& 19.11& -0.41&  0.45& 1.601& -25.74& 1& 0& 0\\
 21& 25& 46.60&  -44& 32& 58.3& 20.39& -0.04&  0.56& 2.503& -25.39& 1& 0& 0\\
 21& 25& 53.43&  -43& 43& 39.9& 20.44& -0.29&  0.65& 0.451& -21.71& 1& 2& 0\\
 21& 25& 56.64&  -44&  8& 32.1& 18.48& -0.33&  0.63& 0.366& -23.22& 1& 0& 0\\
 21& 25& 49.78&  -46& 11& 32.1& 20.32& -0.78&  1.12& 1.305& -24.10& 1& 2& 3\\
 21& 25& 49.84&  -46& 47& 52.4& 19.07& -0.45&  0.50& 1.888& -26.13& 1& 0& 0\\
 21& 26& 10.02&  -42& 56& 30.8& 17.79& -0.28&  0.67& 0.405& -24.13& 1& 0& 0\\
 21& 26&  1.19&  -47&  8& 26.6& 19.28& -0.10&  0.52& 0.698& -23.81& 1& 0& 0\\
 21& 26&  7.35&  -45& 34& 52.2& 21.27& -0.22&  0.51& 0.719& -21.88& 0& 2& 0\\
 21& 26& 12.68&  -47&  8& 30.4& 20.04&  0.02&  0.68& 2.200& -25.47& 1& 0& 0\\
 21& 26& 33.72&  -45& 58& 45.3& 18.17& -0.23&  0.80& 1.579& -26.65& 1& 2& 0\\
 21& 26& 40.88&  -43& 34& 19.4& 18.48& -0.27&  0.86& 0.584& -24.23& 1& 2& 0\\
 21& 26& 43.20&  -43& 36& 13.8& 20.71& -0.49&  0.68& 1.280& -23.67& 1& 2& 0\\
 21& 26& 39.70&  -45&  2& 43.9& 20.54& -0.49&  0.77& 2.156& -24.93& 0& 2& 0\\
 21& 26& 41.19&  -45& 31& 12.6& 19.93& -0.56&  0.71& 0.980& -23.89& 1& 2& 0\\
 21& 26& 45.68&  -43& 50&  6.3& 20.08& -0.37&  1.14& 1.120& -24.02& 1& 2& 3\\
 21& 26& 50.42&  -43& 46& 50.6& 20.29& -0.47&  1.05& 1.110& -23.79& 1& 2& 0\\
 21& 26& 49.67&  -45& 26& 56.3& 18.59& -0.47&  0.44& 1.382& -25.95& 1& 2& 0\\
 21& 26& 52.97&  -46& 19&  0.7& 18.90& -0.66&  0.67& 1.880& -26.29& 1& 0& 0\\
 21& 27&  0.30&  -44& 56& 34.0& 20.33& -0.16&  1.94& 0.522& -22.14& 0& 0& 3\\
 21& 26& 58.26&  -47&  2& 58.9& 20.69&  0.08&  0.90& 2.592& -25.16& 1& 0& 0\\
 21& 27&  8.64&  -44& 29& 24.6& 20.83& -0.20&  0.90& 2.122& -24.61& 0& 2& 0\\
 21& 27& 13.56&  -44& 35& 17.7& 18.82& -0.57&  0.69& 2.015& -26.51& 1& 2& 0\\
 21& 27& 19.33&  -42& 42& 44.3& 17.71& -0.41&  0.66& 0.799& -25.67& 1& 0& 0\\
 21& 27& 14.13&  -45& 55& 19.9& 20.12&  0.15&  0.66& 2.440& -25.61& 1& 0& 0\\
 21& 27& 19.62&  -43& 45& 11.4& 20.08& -0.46&  0.72& 1.722& -24.92& 1& 2& 0\\
 21& 27& 17.65&  -46&  1& 39.1& 19.98& -0.94&  1.37& 1.285& -24.41& 1& 2& 3\\
\end{tabular}

\end{table*}

\begin{table*}[t]

\begin{tabular}[b]{rrrrrrrrrrrrrr}
\multicolumn{3}{c}{R.A. (1950)}&\multicolumn{3}{c}{Dec (1950)}&
\multicolumn{1}{c}{$B$}&\multicolumn{1}{c}{$U\!-\!B$}&
\multicolumn{1}{c}{$\delta m$}& \multicolumn{1}{c}{$z$}&
\multicolumn{1}{c}{$M_{B}$}&\multicolumn{3}{c}{Samples}\\
 &&&&&&&&&&&&&\\
 21& 27& 21.35&  -44& 56& 24.6& 19.83&  0.43&  0.63& 2.880& -26.24& 1& 0& 0\\
 21& 27& 31.32&  -44& 45&  6.4& 18.72& -0.35&  0.78& 1.536& -26.04& 1& 0& 0\\
 21& 27& 39.34&  -43&  8& 36.7& 20.46&  0.00&  0.84& 2.674& -25.45& 1& 0& 0\\
 21& 27& 37.22&  -45& 28& 57.0& 18.60&  0.84&  0.44& 2.730& -27.36& 1& 0& 0\\
 21& 27& 52.99&  -43& 50& 37.0& 19.40& -0.15&  0.64& 0.493& -22.94& 1& 0& 0\\
 21& 27& 54.41&  -45& 52& 55.5& 19.94& -0.48&  1.17& 1.101& -24.12& 0& 0& 3\\
 21& 27& 55.93&  -45& 49&  0.5& 19.04& -0.15&  1.07& 0.871& -24.52& 1& 0& 0\\
 21& 27& 55.24&  -47&  6& 50.1& 18.42& -0.24&  0.57& 0.578& -24.27& 1& 0& 0\\
 21& 27& 59.09&  -45& 11&  9.3& 19.83& -0.89&  0.83& 2.001& -25.49& 1& 2& 0\\
 21& 28&  2.17&  -45& 25& 10.3& 19.01& -0.37&  0.59& 0.961& -24.76& 0& 2& 0\\
 21& 28&  4.28&  -46& 39&  7.9& 20.73&  0.05&  0.71& 2.610& -25.13& 1& 0& 0\\
 21& 28&  5.82&  -46& 33& 24.7& 17.44&  0.19&  0.65& 0.153& -22.37& 1& 0& 0\\
 21& 28& 14.96&  -42& 45& 41.9& 19.44& -0.38&  0.64& 1.420& -25.16& 1& 0& 0\\
 21& 28& 13.84&  -44& 43& 32.1& 20.49& -0.23&  0.62& 1.229& -23.81& 1& 2& 0\\
 21& 28& 17.05&  -44& 34& 56.0& 19.52& -0.53&  0.68& 1.434& -25.10& 1& 2& 0\\
 21& 28& 21.96&  -46& 10& 53.2& 17.65& -0.50&  0.40& 0.833& -25.82& 1& 2& 0\\
 21& 28& 24.69&  -45& 37&  3.4& 19.77&  0.00&  0.70& 0.204& -20.67& 1& 0& 0\\
 21& 28& 40.98&  -45& 20& 20.8& 20.56& -0.25&  1.75& 0.544& -22.00& 0& 0& 3\\
 21& 28& 42.00&  -43& 30& 59.2& 18.91& -0.60&  0.42& 1.905& -26.30& 1& 0& 0\\
 21& 28& 44.77&  -46&  4& 16.3& 20.51& -0.20&  1.12& 0.831& -22.95& 1& 2& 0\\
 21& 28& 56.74&  -45& 48&  1.9& 20.45& -0.42&  0.98& 1.087& -23.59& 0& 2& 0\\
 21& 29&  1.26&  -46&  6& 30.4& 20.07& -0.40&  0.81& 1.025& -23.84& 1& 2& 0\\
 21& 29&  4.08&  -42& 55& 55.0& 20.65& -0.46&  1.10& 2.070& -24.74& 0& 0& 3\\
 21& 29&  4.72&  -44& 42& 39.7& 19.22& -0.38&  0.99& 1.025& -24.69& 1& 2& 0\\
 21& 29&  8.42&  -45& 15& 39.0& 19.80& -0.67&  0.61& 1.260& -24.55& 1& 2& 0\\
 21& 29& 10.18&  -46& 17&  2.3& 20.74& -0.30&  1.31& 0.998& -23.11& 1& 2& 0\\
 21& 29& 10.53&  -44& 50& 18.5& 19.07& -0.68&  1.03& 1.942& -26.18& 1& 2& 0\\
 21& 29& 24.73&  -45& 27&  7.0& 20.21& -0.66&  0.84& 1.375& -24.32& 1& 2& 0\\
 21& 29& 27.63&  -45&  1& 15.8& 20.89& -0.08&  0.56& 2.170& -24.59& 1& 0& 0\\
 21& 29& 32.92&  -45& 51& 30.9& 20.58& -0.35&  0.82& 1.286& -23.81& 1& 2& 0\\
 21& 29& 38.02&  -45& 12& 11.6& 19.31&  0.11&  0.90& 2.180& -26.18& 1& 0& 0\\
 21& 29& 38.26&  -44& 13& 47.1& 19.85& -0.45&  0.83& 1.452& -24.80& 1& 2& 0\\
 21& 29& 39.50&  -46& 24& 19.2& 17.80& -0.30&  0.67& 0.435& -24.27& 1& 0& 0\\
 21& 29& 39.63&  -46& 29& 40.3& 19.71&  0.02&  0.65& 2.465& -26.04& 1& 0& 0\\
 21& 29& 41.25&  -45& 22& 49.0& 20.21&  9.99&  0.48& 3.580& -26.30& 1& 0& 0\\
 21& 29& 42.67&  -45& 16& 27.2& 20.62& -0.54&  0.73& 1.030& -23.30& 1& 2& 0\\
 21& 29& 45.45&  -46& 53& 47.6& 18.85& -0.02&  0.61& 2.230& -26.69& 1& 0& 0\\
 21& 29& 47.37&  -46&  2&  7.8& 18.97& -0.27&  0.51& 0.299& -22.29& 1& 2& 0\\
 21& 29& 50.05&  -45& 46& 56.6& 20.29& -0.25&  0.62& 2.034& -25.06& 0& 2& 0\\
 21& 29& 51.34&  -44& 25&  9.8& 20.71& -0.19&  0.86& 0.744& -22.52& 1& 0& 0\\
 21& 29& 57.55&  -46& 53& 51.1& 19.85&  0.02&  0.62& 2.208& -25.67& 1& 0& 0\\
 21& 29& 56.13&  -44& 23& 10.5& 20.76& -0.67&  0.78& 2.011& -24.57& 0& 2& 0\\
 21& 30&  1.27&  -46&  3&  4.9& 19.61& -0.14&  0.52& 2.244& -25.94& 1& 0& 0\\
 21& 30&  1.99&  -44&  3& 36.7& 20.88&  0.68&  0.74& 2.970& -25.25& 1& 0& 0\\
 21& 30&  2.98&  -45&  8& 45.2& 19.07&  0.05&  1.38& 0.740& -24.15& 1& 0& 3\\
 21& 30&  8.22&  -43& 52& 59.0& 19.93&  0.11&  0.84& 2.634& -25.95& 1& 0& 0\\
 21& 30& 16.27&  -43& 10& 28.9& 21.37& -0.16&  1.41& 0.914& -22.30& 0& 0& 3\\
 21& 30& 20.02&  -44& 40& 42.8& 20.25& -0.07&  0.47& 0.610& -22.55& 1& 0& 0\\
 21& 30& 20.78&  -44& 45& 36.3& 20.00& -0.48&  0.59& 1.460& -24.66& 0& 2& 0\\
 21& 30& 22.37&  -44& 50& 58.9& 19.65& -0.12&  0.61& 0.725& -23.52& 1& 0& 0\\
 21& 30& 24.58&  -44& 42& 19.8& 20.29&  0.96&  0.40& 3.040& -25.89& 1& 0& 0\\
 21& 30& 27.93&  -45& 55& 32.9& 18.91& -0.31&  0.40& 1.556& -25.88& 1& 2& 0\\
 21& 30& 30.00&  -45& 45& 38.2& 20.65&  0.23&  0.59& 2.645& -25.24& 1& 0& 0\\
 21& 30& 30.16&  -44& 45& 24.4& 19.71& -0.63&  0.64& 1.324& -24.74& 0& 2& 0\\
 21& 30& 30.63&  -43&  6& 51.6& 19.51& -0.23&  0.69& 1.645& -25.40& 1& 2& 0\\
\end{tabular}

\end{table*}

\begin{table*}[t]

\begin{tabular}[b]{rrrrrrrrrrrrrr}
\multicolumn{3}{c}{R.A. (1950)}&\multicolumn{3}{c}{Dec (1950)}&
\multicolumn{1}{c}{$B$}&\multicolumn{1}{c}{$U\!-\!B$}&
\multicolumn{1}{c}{$\delta m$}& \multicolumn{1}{c}{$z$}&
\multicolumn{1}{c}{$M_{B}$}&\multicolumn{3}{c}{Samples}\\
 &&&&&&&&&&&&&\\
 21& 30& 38.92&  -43& 50& 22.4& 20.22& -0.69&  1.31& 1.229& -24.08& 1& 2& 3\\
 21& 30& 41.40&  -43& 48&  1.0& 19.56& -0.40&  0.61& 1.597& -25.29& 1& 2& 0\\
 21& 30& 45.86&  -45& 56& 38.6& 20.58& -0.26&  0.61& 1.498& -24.13& 1& 2& 0\\
 21& 30& 50.85&  -44& 32&  7.2& 18.85& -0.27&  0.54& 0.793& -24.51& 1& 2& 0\\
 21& 31&  1.26&  -46&  7&  4.0& 20.73& -0.39&  0.41& 2.010& -24.60& 0& 2& 0\\
 21& 31&  0.28&  -44& 36& 23.0& 19.99& -0.36&  0.47& 1.234& -24.31& 1& 2& 0\\
 21& 31&  7.60&  -46& 12& 42.8& 20.71& -0.40&  0.83& 1.978& -24.58& 0& 2& 0\\
 21& 31&  4.84&  -44& 14&  6.4& 20.23& -0.44&  0.62& 0.999& -23.63& 0& 2& 0\\
 21& 31& 12.57&  -46&  0& 12.2& 19.90&  0.10&  0.65& 2.719& -26.05& 1& 0& 0\\
 21& 31& 14.89&  -46& 56&  9.1& 19.26& -0.30&  0.37& 0.980& -24.56& 1& 0& 0\\
 21& 31&  9.87&  -43& 39& 29.7& 20.76& -0.33&  0.65& 1.390& -23.79& 1& 2& 0\\
 21& 31& 15.15&  -45&  7& 35.1& 19.19& -0.73&  0.52& 1.966& -26.09& 1& 2& 0\\
 21& 31& 15.80&  -45&  1& 19.0& 19.67& -0.22&  1.15& 0.418& -22.32& 0& 0& 3\\
 21& 31& 21.94&  -46& 38& 23.6& 18.71& -0.45&  0.54& 1.720& -26.29& 1& 0& 0\\
 21& 31& 15.90&  -42& 43& 18.9& 20.20&  0.08&  0.99& 0.365& -21.50& 1& 0& 0\\
 21& 31& 22.52&  -43& 43& 53.5& 20.55& -0.41&  0.70& 0.844& -22.95& 1& 2& 0\\
 21& 31& 26.19&  -45& 15&  3.1& 20.69& -0.72&  1.12& 1.920& -24.54& 1& 2& 3\\
 21& 31& 27.03&  -45&  4& 49.3& 19.85&  0.00&  1.12& 0.715& -23.29& 1& 0& 0\\
 21& 31& 23.41&  -42& 59& 60.0& 20.13& -0.39&  0.80& 1.648& -24.78& 1& 2& 0\\
 21& 31& 37.75&  -45& 48& 53.2& 20.65& -0.20&  1.37& 0.781& -22.68& 0& 0& 3\\
 21& 31& 35.99&  -44& 10& 45.4& 21.12& -0.62&  1.25& 0.529& -21.38& 0& 2& 3\\
 21& 31& 33.39&  -42& 57& 51.0& 18.22& -0.67&  0.74& 2.096& -27.19& 1& 2& 0\\
 21& 31& 43.86&  -46& 57& 13.0& 17.97&  0.03&  0.54& 0.684& -25.08& 1& 0& 0\\
 21& 31& 36.21&  -44&  3& 24.2& 20.35& -0.36&  0.63& 1.660& -24.58& 0& 2& 0\\
 21& 31& 36.77&  -43& 39& 43.3& 19.90& -0.35&  0.74& 0.843& -23.59& 0& 2& 0\\
 21& 31& 38.88&  -43& 54& 22.2& 20.40& -0.34&  0.42& 1.216& -23.87& 0& 2& 0\\
 21& 31& 38.93&  -42& 47& 29.3& 20.39& -0.36&  1.23& 0.982& -23.43& 1& 2& 3\\
 21& 31& 43.86&  -43& 39& 11.8& 20.52& -0.53&  0.55& 0.715& -22.62& 0& 2& 0\\
 21& 31& 50.98&  -43& 19& 14.5& 20.05& -0.38&  0.51& 1.656& -24.87& 0& 2& 0\\
 21& 31& 54.25&  -44& 29& 23.0& 19.14& -0.49&  0.50& 1.414& -25.45& 1& 0& 0\\
 21& 32&  2.95&  -45& 16&  9.7& 18.91& -0.55&  0.79& 1.890& -26.29& 1& 0& 0\\
 21& 31& 56.70&  -42& 55& 39.3& 21.08& -0.36&  1.70& 1.730& -23.93& 0& 2& 3\\
 21& 32& 10.46&  -46& 14& 20.6& 19.60&  0.34&  0.72& 2.769& -26.39& 1& 0& 0\\
 21& 32&  4.70&  -42& 45& 41.3& 19.02& -0.61&  0.70& 1.992& -26.29& 1& 0& 0\\
 21& 32& 13.49&  -45& 16& 50.0& 17.68& -0.59&  0.55& 0.507& -24.72& 1& 2& 0\\
 21& 32& 13.00&  -44& 14& 14.0& 19.65& -0.22&  0.63& 1.645& -25.26& 1& 2& 0\\
 21& 32& 11.55&  -43& 24& 45.8& 19.87& -0.35&  1.71& 0.860& -23.67& 1& 2& 3\\
 21& 32& 18.25&  -45&  6& 59.8& 19.20& -0.45&  0.61& 0.520& -23.26& 1& 0& 0\\
 21& 32& 18.33&  -44& 18& 48.4& 19.87& -0.35&  0.97& 1.451& -24.77& 1& 2& 0\\
 21& 32& 21.72&  -43& 24& 47.5& 20.43& -0.11&  0.52& 2.240& -25.12& 1& 0& 0\\
 21& 32& 23.85&  -43& 48& 10.8& 20.19&  0.23&  0.86& 2.790& -25.81& 1& 0& 0\\
 21& 32& 23.30&  -43&  9& 42.2& 20.82& -0.56&  0.45& 0.959& -22.95& 0& 2& 0\\
 21& 32& 27.48&  -44&  5& 24.8& 20.41& -0.28&  0.98& 0.478& -21.87& 1& 2& 0\\
 21& 32& 36.44&  -46& 34&  0.5& 19.46& -0.53&  1.33& 1.320& -24.99& 1& 0& 3\\
 21& 32& 35.27&  -46& 11& 31.5& 19.01& -0.46&  0.66& 1.600& -25.84& 1& 2& 0\\
 21& 32& 26.58&  -43& 18&  4.3& 19.66& -0.31&  0.82& 1.710& -25.33& 1& 2& 0\\
 21& 32& 31.07&  -44& 18& 57.9& 19.08& -0.77&  1.25& 1.258& -25.26& 1& 2& 3\\
 21& 32& 37.49&  -45&  8& 31.6& 20.12& -0.56&  0.59& 1.377& -24.41& 1& 2& 0\\
 21& 32& 37.63&  -44& 51&  1.2& 18.55& -0.41&  0.91& 0.920& -25.13& 1& 0& 0\\
 21& 32& 38.00&  -44& 20& 41.8& 20.61& -0.53&  1.12& 0.655& -22.34& 1& 2& 0\\
 21& 32& 54.14&  -47&  5&  8.9& 19.38& -0.44&  1.59& 0.244& -21.44& 1& 0& 3\\
 21& 32& 53.63&  -45& 58& 49.2& 20.29& -0.41&  0.76& 1.003& -23.57& 1& 2& 0\\
 21& 32& 51.16&  -44& 31& 17.9& 21.33& -0.09&  1.76& 0.430& -20.72& 0& 0& 3\\
 21& 33&  3.03&  -46& 20& 24.7& 20.87&  0.27&  0.78& 2.760& -25.11& 1& 0& 0\\
 21& 32& 55.05&  -43& 21& 44.1& 17.93&  0.27&  0.48& 2.420& -27.78& 1& 0& 0\\
\end{tabular}

\end{table*}

\begin{table*}[t]

\begin{tabular}[b]{rrrrrrrrrrrrrr}
\multicolumn{3}{c}{R.A. (1950)}&\multicolumn{3}{c}{Dec (1950)}&
\multicolumn{1}{c}{$B$}&\multicolumn{1}{c}{$U\!-\!B$}&
\multicolumn{1}{c}{$\delta m$}& \multicolumn{1}{c}{$z$}&
\multicolumn{1}{c}{$M_{B}$}&\multicolumn{3}{c}{Samples}\\
 &&&&&&&&&&&&&\\
 21& 33&  2.60&  -44& 16& 33.7& 20.06& -0.29&  0.28& 0.327& -21.40& 0& 2& 0\\
 21& 33&  6.66&  -44& 28& 26.5& 20.80& -0.68&  0.99& 0.544& -21.76& 1& 2& 0\\
 21& 33&  7.60&  -44& 16& 50.7& 20.76&  9.99&  0.75& 3.340& -25.61& 1& 0& 0\\
 21& 33& 13.11&  -45& 20&  2.1& 20.65&  0.34&  0.49& 2.898& -25.43& 1& 0& 0\\
 21& 33&  7.84&  -43&  4& 13.9& 20.25&  0.19&  1.00& 2.470& -25.50& 1& 0& 0\\
 21& 33& 12.34&  -42& 58& 51.0& 19.40& -0.22&  0.43& 1.190& -24.83& 1& 2& 0\\
 21& 33& 22.63&  -44& 20& 23.7& 19.96& -0.37&  0.45& 0.999& -23.90& 1& 2& 0\\
 21& 33& 31.09&  -46& 11& 38.3& 19.72& -0.37&  0.65& 0.888& -23.89& 1& 2& 0\\
 21& 33& 31.37&  -46& 13& 34.8& 19.58& -0.44&  1.36& 1.448& -25.06& 1& 2& 3\\
 21& 33& 21.32&  -43& 30& 55.6& 21.29& -0.55&  0.92& 1.720& -23.71& 0& 2& 0\\
 21& 33& 23.97&  -43& 36& 17.3& 20.42& -0.20&  0.46& 0.669& -22.58& 1& 2& 0\\
 21& 33& 29.69&  -44&  4&  8.7& 20.37& -0.22&  0.90& 1.756& -24.67& 1& 2& 0\\
 21& 33& 39.37&  -46&  3&  3.6& 20.75&  0.03&  1.45& 0.423& -21.26& 1& 0& 3\\
 21& 33& 26.75&  -42& 55&  1.3& 20.45& -0.14&  1.57& 0.580& -22.24& 0& 0& 3\\
 21& 33& 38.98&  -45& 34& 51.6& 19.44& -0.24&  1.54& 0.868& -24.12& 1& 2& 3\\
 21& 33& 41.36&  -45& 53&  0.8& 18.34& -0.54&  0.36& 1.000& -25.52& 1& 2& 0\\
 21& 33& 51.84&  -46& 17& 51.0& 20.75& -0.49&  1.21& 1.571& -24.06& 0& 2& 0\\
 21& 33& 53.98&  -46& 17& 24.2& 19.65& -0.40&  0.85& 0.844& -23.85& 1& 2& 0\\
 21& 33& 58.56&  -47&  8&  7.8& 19.35& -0.62&  1.10& 1.097& -24.70& 1& 0& 3\\
 21& 33& 45.77&  -43& 58& 44.9& 18.27& -0.33&  0.83& 1.560& -26.53& 1& 2& 0\\
 21& 33& 45.15&  -43& 49& 23.8& 20.09& -0.59&  0.77& 1.850& -25.06& 1& 2& 0\\
 21& 33& 47.57&  -44& 17& 35.9& 20.35&  0.44&  0.45& 2.838& -25.69& 1& 0& 0\\
 21& 33& 48.75&  -44& 28& 48.7& 20.21& -0.51&  0.51& 2.059& -25.17& 0& 2& 0\\
 21& 34&  1.71&  -46& 46& 47.6& 19.90&  0.63&  0.74& 3.065& -26.29& 1& 0& 0\\
 21& 33& 53.80&  -44& 56& 27.5& 20.08&  0.03&  0.62& 2.745& -25.89& 1& 0& 0\\
 21& 33& 46.40&  -43& 21& 34.2& 20.24& -0.69&  0.83& 2.022& -25.10& 1& 2& 0\\
 21& 33& 51.02&  -43&  6& 10.0& 20.81& -0.55&  0.85& 1.400& -23.76& 1& 2& 0\\
 21& 34&  8.30&  -45&  6&  3.0& 20.59& -0.31&  0.53& 0.880& -23.00& 1& 2& 0\\
 21& 34&  2.22&  -43& 10& 38.6& 21.12& -0.50&  0.56& 0.524& -21.36& 0& 2& 0\\
 21& 34&  8.13&  -43& 51& 26.6& 19.56& -0.24&  1.03& 1.663& -25.37& 1& 2& 0\\
 21& 34& 19.95&  -46&  2& 17.9& 20.41& -0.65&  1.52& 1.292& -23.99& 1& 2& 3\\
 21& 34&  8.81&  -43& 11& 33.7& 19.92& -0.24&  0.74& 1.650& -24.99& 1& 2& 0\\
 21& 34& 19.94&  -45& 16& 58.6& 20.44& -0.48&  0.84& 1.106& -23.63& 1& 2& 0\\
 21& 34& 17.90&  -43& 50&  8.8& 19.88& -0.35&  0.82& 1.558& -24.91& 1& 2& 0\\
 21& 34& 13.93&  -42& 57& 43.8& 19.30&  0.11&  0.60& 2.608& -26.56& 1& 0& 0\\
 21& 34& 18.64&  -43& 52& 39.7& 20.71& -0.61&  0.64& 1.370& -23.81& 1& 2& 0\\
 21& 34& 37.07&  -46& 45&  4.8& 19.37& -0.23&  0.82& 0.704& -23.74& 1& 0& 0\\
 21& 34& 30.32&  -44& 35& 55.7& 20.27& -0.72&  1.48& 1.319& -24.17& 0& 0& 3\\
 21& 34& 29.90&  -44&  3&  5.5& 20.47& -0.12&  1.11& 1.576& -24.35& 1& 0& 0\\
 21& 34& 47.85&  -46&  2& 39.1& 20.15& -0.52&  0.56& 1.340& -24.33& 0& 2& 0\\
 21& 34& 50.90&  -46&  1& 49.4& 17.81&  0.12&  0.46& 0.528& -24.68& 1& 0& 0\\
 21& 34& 37.41&  -43& 40& 37.4& 19.39& -0.25&  0.48& 0.837& -24.09& 0& 2& 0\\
 21& 34& 49.15&  -44& 43&  0.2& 19.53&  0.15&  0.64& 2.526& -26.27& 1& 0& 0\\
 21& 35&  0.95&  -46& 11& 36.5& 18.87& -0.41&  0.27& 1.903& -26.34& 0& 2& 0\\
 21& 34& 43.15&  -43&  6& 56.5& 19.67& -0.36&  0.81& 1.423& -24.93& 1& 2& 0\\
 21& 35&  6.63&  -46& 32& 27.6& 18.66& -0.08&  1.02& 2.222& -26.87& 1& 0& 0\\
 21& 35&  2.90&  -45& 49& 32.5& 20.06& -0.59&  0.72& 1.832& -25.07& 1& 2& 0\\
 21& 35&  4.13&  -45& 47&  3.5& 20.36& -0.46&  1.15& 1.845& -24.79& 0& 2& 0\\
 21& 35&  0.77&  -44& 16& 28.6& 20.67& -0.27&  0.26& 0.285& -20.49& 0& 2& 0\\
 21& 34& 56.25&  -43& 13& 47.4& 20.08& -0.30&  0.85& 2.141& -25.38& 1& 2& 0\\
 21& 35& 19.47&  -46& 20& 47.4& 18.41& -0.21&  0.53& 0.505& -23.99& 1& 0& 0\\
 21& 35&  0.16&  -43&  4& 14.3& 19.60& -0.26&  1.96& 0.511& -22.82& 1& 2& 3\\
 21& 35&  4.58&  -43& 46& 13.0& 20.87&  0.50&  0.59& 3.100& -25.35& 1& 0& 0\\
 21& 35& 14.93&  -45& 16& 31.0& 20.94& -0.20&  1.21& 0.655& -22.01& 1& 2& 3\\
 21& 35&  5.14&  -43&  3& 48.9& 19.62& -0.45&  0.47& 2.029& -25.72& 1& 2& 0\\
\end{tabular}

\end{table*}

\begin{table*}[t]

\begin{tabular}[b]{rrrrrrrrrrrrrr}
\multicolumn{3}{c}{R.A. (1950)}&\multicolumn{3}{c}{Dec (1950)}&
\multicolumn{1}{c}{$B$}&\multicolumn{1}{c}{$U\!-\!B$}&
\multicolumn{1}{c}{$\delta m$}& \multicolumn{1}{c}{$z$}&
\multicolumn{1}{c}{$M_{B}$}&\multicolumn{3}{c}{Samples}\\
 &&&&&&&&&&&&&\\
 21& 35& 12.41&  -44&  2& 15.0& 19.30& -0.33&  0.34& 0.929& -24.40& 0& 2& 0\\
 21& 35& 28.41&  -45& 59&  2.1& 19.95& -0.56&  0.72& 0.568& -22.70& 1& 2& 0\\
 21& 35& 13.86&  -43& 24& 17.3& 19.18& -0.42&  0.77& 1.935& -26.07& 1& 2& 0\\
 21& 35& 28.59&  -45& 29& 56.1& 20.77& -0.27&  0.64& 2.148& -24.69& 1& 2& 0\\
 21& 35& 28.35&  -45& 23& 57.7& 18.69& -0.47&  0.66& 1.097& -25.36& 1& 2& 0\\
 21& 35& 22.69&  -44&  7& 50.2& 20.89& -0.36&  0.75& 1.168& -23.30& 1& 2& 0\\
 21& 35& 20.13&  -42& 53& 29.7& 17.89& -0.45&  0.64& 1.469& -26.78& 1& 2& 0\\
 21& 35& 23.26&  -43& 20&  7.1& 20.33& -0.33&  1.00& 1.535& -24.43& 1& 2& 0\\
 21& 35& 34.64&  -43& 49& 30.6& 21.12& -0.54&  0.86& 1.395& -23.44& 0& 2& 0\\
 21& 35& 41.40&  -44&  0& 49.2& 18.45& -0.32&  0.87& 0.461& -23.75& 0& 2& 0\\
 21& 35& 38.64&  -43& 26& 25.7& 20.65&  0.06&  0.46& 2.483& -25.11& 1& 0& 0\\
 21& 35& 43.33&  -44&  8& 11.1& 20.59& -0.48&  1.23& 1.785& -24.49& 0& 2& 0\\
 21& 35& 47.12&  -44&  1& 23.2& 20.41& -0.18&  0.60& 1.708& -24.58& 1& 0& 0\\
 21& 35& 43.26&  -43& 20& 54.9& 20.38& -0.11&  0.55& 2.186& -25.12& 1& 0& 0\\
 21& 36&  6.35&  -46& 26&  0.5& 19.48& -0.37&  1.31& 0.920& -24.20& 0& 0& 3\\
 21& 35& 53.01&  -44& 21& 44.3& 18.69& -0.15&  0.70& 0.408& -23.25& 1& 0& 0\\
 21& 35& 54.17&  -44& 14& 32.4& 21.14& -0.51&  0.27& 0.412& -20.82& 0& 2& 0\\
 21& 35& 54.42&  -43& 45& 24.2& 20.47& -0.43&  0.98& 1.633& -24.42& 1& 2& 0\\
 21& 36& 19.27&  -46& 48& 48.0& 18.88& -0.11&  0.54& 1.124& -25.23& 1& 0& 0\\
 21& 36& 17.27&  -44& 30& 11.6& 19.80& -0.46&  0.41& 0.931& -23.91& 0& 2& 0\\
 21& 36& 28.18&  -45& 38&  4.6& 19.14& -0.22&  0.59& 1.695& -25.83& 1& 0& 0\\
 21& 36& 21.01&  -43& 17& 51.2& 20.29& -0.39&  0.98& 1.048& -23.67& 0& 2& 0\\
 21& 36& 31.16&  -44& 20& 53.9& 21.14& -0.35&  0.29& 0.433& -20.92& 0& 2& 0\\
 21& 36& 28.71&  -43& 47& 29.6& 18.96& -0.34&  0.61& 1.520& -25.78& 1& 2& 0\\
 21& 36& 30.45&  -43&  1& 22.7& 18.12& -0.57&  0.62& 1.343& -26.36& 1& 0& 0\\
 21& 36& 39.18&  -44& 11& 35.1& 19.63& -0.41&  0.69& 1.925& -25.61& 1& 2& 0\\
 21& 36& 52.42&  -45& 29&  1.7& 19.43& -0.09&  0.75& 1.174& -24.77& 1& 0& 0\\
 21& 36& 34.83&  -42& 48&  0.0& 20.10&  0.08&  0.40& 0.168& -19.92& 1& 0& 0\\
 21& 37&  4.86&  -46& 20& 11.6& 19.22& -0.02&  0.52& 2.495& -26.55& 1& 0& 0\\
 21& 36& 42.71&  -43& 30& 28.1& 20.86& -0.33&  1.41& 1.028& -23.06& 1& 2& 3\\
 21& 36& 49.17&  -44& 20& 47.4& 20.23& -0.30&  0.35& 0.275& -20.85& 0& 2& 0\\
 21& 36& 42.52&  -43& 28& 46.6& 18.38& -0.43&  0.54& 0.250& -22.50& 1& 2& 0\\
 21& 36& 59.31&  -45& 17& 26.5& 20.67& -0.38&  0.86& 1.105& -23.40& 1& 2& 0\\
 21& 36& 44.29&  -42& 51& 36.2& 19.67& -0.41&  0.78& 1.510& -25.06& 1& 2& 0\\
 21& 37&  8.20&  -45& 23& 36.4& 20.45& -0.38&  0.94& 1.645& -24.46& 1& 2& 0\\
 21& 37&  2.10&  -44& 22& 56.2& 20.64& -0.52&  0.75& 2.050& -24.73& 0& 2& 0\\
 21& 37& 12.13&  -45& 20& 33.6& 20.70& -0.34&  0.58& 2.113& -24.73& 0& 2& 0\\
 21& 37&  6.81&  -43& 54& 27.7& 19.70& -0.77&  0.75& 1.915& -25.52& 1& 2& 0\\
 21& 37& 25.87&  -45& 29& 27.9& 20.33& -0.39&  0.60& 1.438& -24.30& 1& 2& 0\\
 21& 37& 14.45&  -44&  6& 39.4& 20.30& -0.05&  0.62& 2.250& -25.26& 1& 0& 0\\
 21& 37& 17.79&  -44& 21& 10.6& 19.72& -0.35&  1.42& 0.346& -21.86& 1& 2& 3\\
 21& 37& 17.76&  -44& 10& 44.8& 20.87& -0.42&  0.65& 0.798& -22.51& 0& 2& 0\\
 21& 37& 10.59&  -43&  5& 26.6& 20.92& -0.42&  1.32& 1.393& -23.64& 1& 2& 3\\
 21& 37& 29.30&  -44& 49& 49.2& 19.54&  0.49&  0.72& 0.136& -20.02& 1& 0& 0\\
 21& 37& 15.41&  -43&  1& 47.3& 19.83& -0.22&  1.11& 1.520& -24.91& 1& 2& 3\\
 21& 37& 31.36&  -44& 22& 18.3& 19.42&  0.32&  0.54& 1.320& -25.03& 1& 0& 0\\
 21& 37& 53.50&  -46& 45& 21.6& 20.42&  0.24&  1.11& 2.800& -25.59& 1& 0& 0\\
 21& 37& 32.47&  -43& 36& 56.0& 19.53& -0.10&  0.48& 2.310& -26.08& 1& 0& 0\\
 21& 37& 38.02&  -43& 43&  5.3& 19.16& -0.74&  0.50& 1.964& -26.12& 1& 2& 0\\
 21& 38&  0.90&  -46& 11& 10.3& 19.48&  0.04&  0.66& 2.287& -26.11& 1& 0& 0\\
 21& 37& 51.68&  -44& 35& 48.8& 18.34& -0.34&  0.35& 0.632& -24.54& 1& 0& 0\\
 21& 37& 45.95&  -43& 56& 50.9& 20.02& -0.63&  1.67& 2.039& -25.34& 0& 2& 3\\
 21& 38& 17.41&  -46& 35&  7.3& 21.02& -0.37&  1.40& 0.780& -22.31& 0& 0& 3\\
 21& 38& 15.98&  -46& 21& 31.8& 19.45& -0.16&  0.61& 0.618& -23.38& 1& 0& 0\\
 21& 38& 16.48&  -45& 12& 28.1& 19.34& -0.24&  0.25& 2.164& -26.14& 0& 2& 0\\
\end{tabular}

\end{table*}

\begin{table*}[t]

\begin{tabular}[b]{rrrrrrrrrrrrrr}
\multicolumn{3}{c}{R.A. (1950)}&\multicolumn{3}{c}{Dec (1950)}&
\multicolumn{1}{c}{$B$}&\multicolumn{1}{c}{$U\!-\!B$}&
\multicolumn{1}{c}{$\delta m$}& \multicolumn{1}{c}{$z$}&
\multicolumn{1}{c}{$M_{B}$}&\multicolumn{3}{c}{Samples}\\
 &&&&&&&&&&&&&\\
 21& 38& 11.93&  -44& 20& 28.7& 19.66& -0.59&  0.51& 2.102& -25.76& 1& 2& 0\\
 21& 38&  5.54&  -43& 20& 26.1& 19.91& -0.17&  0.54& 0.574& -22.76& 1& 0& 0\\
 21& 38& 41.58&  -46& 50& 33.2& 18.26& -0.20&  1.02& 0.762& -25.02& 1& 0& 0\\
 21& 38& 14.85&  -43& 45& 49.7& 19.26& -0.48&  0.93& 1.320& -25.19& 1& 2& 0\\
 21& 38& 12.87&  -43&  9& 10.4& 20.22& -0.17&  2.08& 0.687& -22.84& 1& 0& 3\\
 21& 38& 25.45&  -44&  5& 15.3& 19.85& -0.31&  0.73& 0.862& -23.69& 1& 2& 0\\
 21& 38& 15.73&  -42& 55& 44.8& 20.13& -0.23&  0.83& 1.597& -24.72& 1& 2& 0\\
 21& 38& 33.14&  -44& 16& 15.2& 20.53& -0.33&  1.73& 0.623& -22.32& 0& 0& 3\\
 21& 38& 28.59&  -43&  7&  4.9& 20.50& -0.30&  0.26& 0.255& -20.42& 0& 2& 0\\
 21& 38& 43.40&  -44& 32& 17.9&  0.00& -0.51&  1.44& 1.660& -44.93& 0& 0& 3\\
 21& 38& 31.80&  -43& 13& 51.2& 20.06& -0.75&  0.79& 1.287& -24.33& 1& 2& 0\\
 21& 38& 56.08&  -45& 26&  4.1& 20.08& -0.51&  0.36& 1.197& -24.16& 1& 2& 0\\
 21& 38& 56.93&  -45& 16& 15.5& 20.50& -0.50&  1.50& 2.013& -24.83& 0& 2& 3\\
 21& 39&  0.08&  -45& 20& 38.8& 20.01& -0.20&  0.64& 2.210& -25.51& 1& 2& 0\\
 21& 38& 53.72&  -44& 38& 39.4& 19.50& -0.53&  0.61& 1.078& -24.52& 1& 0& 0\\
 21& 39& 12.24&  -46& 14& 18.8& 20.15&  1.14&  0.59& 3.340& -26.22& 1& 0& 0\\
 21& 38& 57.01&  -44& 16& 52.4& 19.31& -0.37&  0.62& 1.735& -25.71& 1& 2& 0\\
 21& 38& 54.54&  -43& 41& 29.9& 20.74& -0.33&  1.02& 2.210& -24.78& 1& 2& 0\\
 21& 38& 59.14&  -43& 50& 57.6& 18.71&  0.21&  0.40& 0.142& -20.94& 1& 0& 0\\
 21& 38& 52.05&  -42& 58& 42.7& 20.26& -0.46&  0.56& 0.937& -23.46& 1& 2& 0\\
 21& 38& 54.79&  -42& 55& 28.0& 19.87& -0.20&  1.18& 0.542& -22.68& 1& 2& 3\\
 21& 39& 25.50&  -45& 20& 14.0& 19.97& -0.69&  0.79& 1.394& -24.59& 1& 2& 0\\
 21& 39& 20.18&  -44& 43& 60.0& 20.96&  0.09&  0.77& 2.380& -24.71& 1& 0& 0\\
 21& 39& 27.52&  -45& 21& 30.5& 20.73& -0.24&  0.79& 1.671& -24.21& 1& 2& 0\\
 21& 39&  9.07&  -42& 54& 48.2& 20.62& -0.32&  0.73& 1.905& -24.59& 0& 2& 0\\
 21& 39& 22.45&  -43& 32&  4.6& 18.40&  0.11&  0.61& 2.190& -27.10& 1& 0& 0\\
 21& 39& 32.94&  -43& 55& 45.4& 19.47& -0.51&  0.64& 0.679& -23.56& 1& 2& 0\\
 21& 39& 44.37&  -44& 49& 37.2& 20.10& -0.14&  1.46& 1.640& -24.80& 0& 0& 3\\
 21& 39& 27.24&  -43&  8& 28.0& 20.86& -0.20&  1.04& 0.752& -22.39& 1& 2& 0\\
 21& 40&  3.43&  -46& 20&  2.2& 18.78& -0.22&  0.65& 0.548& -23.79& 1& 0& 0\\
 21& 39& 30.84&  -42& 59& 26.0& 19.81& -0.35&  0.59& 1.890& -25.39& 1& 2& 0\\
 21& 39& 42.04&  -43& 28& 17.4& 20.55& -0.29&  0.85& 0.708& -22.57& 0& 2& 0\\
 21& 39& 48.87&  -43& 36& 18.4& 20.86& -0.56&  1.21& 1.379& -23.68& 0& 0& 3\\
 21& 40& 16.52&  -45& 52& 37.0& 18.11& -0.34&  0.71& 1.688& -26.85& 1& 0& 0\\
 21& 39& 51.98&  -43& 14& 14.2& 21.21& -0.15&  1.48& 1.645& -23.70& 0& 0& 3\\
\end{tabular}

\end{table*}

\end{document}